\documentclass[aps,prx,twocolumn,superscriptaddress,showpacs,floatfix]{revtex4-1}
\usepackage{amsmath,amssymb,amsfonts,amsthm}
\usepackage{graphicx}
\usepackage{bm}
\usepackage{bbm}
\usepackage{color}
\usepackage{dcolumn}   
\usepackage{epstopdf}
\usepackage[caption=false]{subfig}
\usepackage{xr}
\usepackage{color}
\usepackage{bbold}
\usepackage[colorlinks=true,linkcolor=blue,urlcolor=blue,citecolor=blue]{hyperref}
\usepackage{comment}

\begin{document}

 \newcommand{\breite}{1.0} 

\newtheorem{prop}{Proposition}
\newtheorem{cor}{Corollary} 

\newcommand{\be}{\begin{equation}}
\newcommand{\ee}{\end{equation}}

\newcommand{\bea}{\begin{eqnarray}}
\newcommand{\eea}{\end{eqnarray}}
\newcommand{\lt}{<}
\newcommand{\gt}{>} 

\newcommand{\Reals}{\mathbb{R}}     
\newcommand{\Com}{\mathbb{C}}       
\newcommand{\Nat}{\mathbb{N}}       

\newcommand{\id}{\mathbboldsymbol{1}}    

\newcommand{\Real}{\mathop{\mathrm{Re}}}
\newcommand{\Imag}{\mathop{\mathrm{Im}}}

\def\O{\mbox{$\mathcal{O}$}}   
\def\F{\mathcal{F}}			
\def\sgn{\text{sgn}}

\newcommand{\deo}{\ensuremath{\Delta_0}}
\newcommand{\dea}{\ensuremath{\Delta}}
\newcommand{\ak}{\ensuremath{a_k}}
\newcommand{\ad}{\ensuremath{a^{\dagger}_{-k}}}
\newcommand{\sx}{\ensuremath{\sigma_x}}
\newcommand{\sz}{\ensuremath{\sigma_z}}
\newcommand{\spl}{\ensuremath{\sigma_{+}}}
\newcommand{\smi}{\ensuremath{\sigma_{-}}}
\newcommand{\alk}{\ensuremath{\alpha_{k}}}
\newcommand{\bk}{\ensuremath{\beta_{k}}}
\newcommand{\ok}{\ensuremath{\omega_{k}}}
\newcommand{\vd}{\ensuremath{V^{\dagger}_1}}
\newcommand{\vi}{\ensuremath{V_1}}
\newcommand{\vo}{\ensuremath{V_o}}
\newcommand{\zc}{\ensuremath{\frac{E_z}{E}}}
\newcommand{\xc}{\ensuremath{\frac{\Delta}{E}}}
\newcommand{\xd}{\ensuremath{X^{\dagger}}}
\newcommand{\aok}{\ensuremath{\frac{\alk}{\ok}}}
\newcommand{\tpw}{\ensuremath{e^{i \ok s }}}
\newcommand{\tpe}{\ensuremath{e^{2iE s }}}
\newcommand{\tmw}{\ensuremath{e^{-i \ok s }}}
\newcommand{\tme}{\ensuremath{e^{-2iE s }}}
\newcommand{\epls}{\ensuremath{e^{F(s)}}}
\newcommand{\emis}{\ensuremath{e^{-F(s)}}}
\newcommand{\epl}{\ensuremath{e^{F(0)}}}
\newcommand{\emi}{\ensuremath{e^{F(0)}}}

\newcommand{\lr}[1]{\left( #1 \right)}
\newcommand{\lrs}[1]{\left( #1 \right)^2}
\newcommand{\lrb}[1]{\left< #1\right>}
\newcommand{\nbt}{\ensuremath{\lr{ \lr{n_k + 1} \tmw + n_k \tpw  }}}

\newcommand{\om}{\ensuremath{\omega}}
\newcommand{\dw}{\ensuremath{\Delta_0}}
\newcommand{\wbp}{\ensuremath{\omega_0}}
\newcommand{\dv}{\ensuremath{\Delta_0}}
\newcommand{\vbp}{\ensuremath{\nu_0}}
\newcommand{\vplus}{\ensuremath{\nu_{+}}}
\newcommand{\vminus}{\ensuremath{\nu_{-}}}
\newcommand{\wplus}{\ensuremath{\omega_{+}}}
\newcommand{\wminus}{\ensuremath{\omega_{-}}}
\newcommand{\uv}[1]{\ensuremath{\mathbf{\hat{#1}}}} 
\newcommand{\abs}[1]{\left| #1 \right|} 
\newcommand{\avg}[1]{\left< #1 \right>} 
\let\underdot=\d 
\renewcommand{\d}[2]{\frac{d #1}{d #2}} 
\newcommand{\dd}[2]{\frac{d^2 #1}{d #2^2}} 
\newcommand{\pd}[2]{\frac{\partial #1}{\partial #2}} 
\newcommand{\pdd}[2]{\frac{\partial^2 #1}{\partial #2^2}} 
\newcommand{\pdc}[3]{\left( \frac{\partial #1}{\partial #2}
 \right)_{#3}} 
\newcommand{\ket}[1]{\left| #1 \right>} 
\newcommand{\bra}[1]{\left< #1 \right|} 
\newcommand{\braket}[2]{\left< #1 \vphantom{#2} \right|
 \left. #2 \vphantom{#1} \right>} 
\newcommand{\matrixel}[3]{\left< #1 \vphantom{#2#3} \right|
 #2 \left| #3 \vphantom{#1#2} \right>} 
\newcommand{\grad}[1]{{\nabla} {#1}} 
\let\divsymb=\div 
\renewcommand{\div}[1]{{\nabla} \cdot \boldsymbol{#1}} 
\newcommand{\curl}[1]{{\nabla} \times \boldsymbol{#1}} 
\newcommand{\laplace}[1]{\nabla^2 \boldsymbol{#1}}
\newcommand{\vs}[1]{\boldsymbol{#1}}
\let\baraccent=\= 

\title{Dynamical approach to improving Majorana qubits and distinguishing them from trivial bound states}

\author{Brett Min}
\email{brettm0320@gmail.com}
\affiliation{Department of Physics, McGill University, Montr\'{e}al, Qu\'{e}bec H3A 2T8, Canada}
\author{Bastien Fajardo}
\affiliation{Department of Physics, McGill University, Montr\'{e}al, Qu\'{e}bec H3A 2T8, Canada}
\author{T.~Pereg-Barnea}
\affiliation{Department of Physics, McGill University, Montr\'{e}al, Qu\'{e}bec H3A 2T8, Canada}
\author{Kartiek Agarwal}
\email{agarwal@physics.mcgill.ca}
\affiliation{Department of Physics, McGill University, Montr\'{e}al, Qu\'{e}bec H3A 2T8, Canada}

\date{\today}
\begin{abstract}
We study a series of dynamical protocols which involve periodically driving a quantum dot coupled to a putative nanowire hosting Majorana zero modes (MZMs) to i) reduce the hybridization between MZMs, ii) improve the coherence of the Majorana qubit with respect to $1/f$ dephasing noise and quasiparticle poisoning, and iii) provide a definitive test to differentiate Andreev Bound states (ABSs) from MZMs. The protocols are based on the notion of {\it draiding} - exchanging a pair of Majoranas twice, repeatedly, at high frequency~\cite{agarwalmajoranadraiding}.  In this process, the exchanged Majorana operators acquire a robust minus sign such that terms in the Hamiltonian, linear in either operator, vanish on average. The four protocols proposed implement draiding by coupling quantum dot(s) to the end(s) of the nanowire.  They are treated using Floquet theory and numerical simulations. The hybridization energy and decoherence rate are shown to be reduced by several orders of magnitude, in accordance with theoretical expectations, when the protocols are implemented on nanowires described by experimentally relevant parameters. The tunneling conductance computed in this Floquet setting reveals zero bias peaks (ZBPs) that become more centered at zero voltage bias. When these protocols are implemented on nanowires supporting ABSs that mimic MZMs in ZBP measurements, the qubit coherence \emph{deteriorates}, in stark contrast to the case where the nanowire supports MZMs and the coherence drastically improves, thus serving as a dynamical test to distinguish MZMs from trivial bound states. 
\end{abstract}
\maketitle




\section{Introduction}

An important platform for building quantum computers that are fault-tolerant at the hardware level is based on Majorana zero modes~\cite{kitaev2001unpaired} (MZMs). These non-Abelian excitations encode the ground state degeneracy of macroscopic quantum systems, with each pair of spatially \emph{isolated} MZMs representing a degeneracy of $2$. The presence of a large number of such MZMs can be used to realize a quantum register whose degeneracy is topologically protected, and is thus largely immune to external noise~\cite{kitaev2001unpaired,sarma2015majorana}. Further, braiding MZMs can be used to realize quantum gates on the register in a topologically protected manner~\cite{ivanov_majorana}. The immense potential of MZMs has resulted in a flurry of theoretical proposals that describe various semiconducting and magnetic elements proximitized with superconductivity that may realize such MZMs~\cite{lutchysausarma,oregrefaelvonoppen,potter2011majorana,alicea2012new,pientka2017topological,fu2008superconducting,lindner2012fractionalizing,lesser2021three}. Many of these platforms have also been investigated experimentally~\cite{mourik2012signatures,deng2012anomalous,das2012zero,churchill2013superconductor,lee2014spin,finck2013anomalous,nadj2014observation,albrecht2016exponential,deng2016majorana,lv2017experimental,zhang2018quantized,vaitiekenas2018effective,deng2018nonlocality,de2018electric,suominen2017zero,nichele2017scaling,ren2019topological,manna2020signature,vaitiekenas2020flux}. 

One of the key challenges that is endemic to nearly all platforms (though exceptions exist) is the requirement of large magnetic fields/moments to coexist with superconducting correlations, phenomena that are antithetical to one another. This makes the experimental realization of MZMs more difficult than previously envisioned. Large magnetic fields, along with disorder, can in some cases also suppress superconductivity completely, giving rise to trivial subgap states such as Andreev bound states (ABSs), that can be remarkably robust to external changes and appear at energies arbitrarily close to zero energy, making them hard to distinguish from MZMs in tunneling measurements~\cite{kells2012near,prada2012transport,san2016majorana,liu2017andreev,moore2018two,setiawan2017electron,aseev2018lifetime,moore2018quantized,liu2018distinguishing,vuik2019reproducing,penaranda2018quantifying,reeg2018zero,stanescu2019robust,pan2020physical}. A more optimistic interpretation is that experiments realize imperfect Majoranas that have significant spatial overlap, and thus hybridize. Two overlapping Majoranas encode a finite energy bound state, which nevertheless can be quite close to zero energy. In summary, two important challenges need to be addressed---i) to devise experiments that can definitively distinguish between MZMs and trivial subgap states such as ABSs, and ii) to further improve MZMs realized under the experimental constraints imposed by material challenges and the required truce between superconducting and magnetic correlations.  

In this work, we attempt to provide answers to these two questions tailored to the quantum wire apparatus in which experimental progress on realizing and detecting MZMs is fairly advanced. We utilize a remarkable property of Majoranas, as first noted in Ref.~\cite{agarwalmajoranadraiding}, that double braiding a pair of Majoranas, $\gamma_L, \gamma_R$, yields a topologically robust minus sign on both the Majorana operators. Periodically performing such double braiding or {\it draiding} can be used in accordance with dynamical decoupling techniques to reduce the hybridization between MZMs $\gamma_{L (R)}$ on the left (right) edge of the quantum wire. 

In particular, inspired by Ref.~\cite{agarwalmajoranadraiding}, we consider coupling such a quantum wire hosting MZMs to one or more quantum dots which in turn are driven periodically to achieve back and forth tunneling of an electron between the quantum dot and the quantum wire. In the adiabatic regime, as determined by the Landau-Zener transition criterion, these transitions can be achieved with efficacy exponentially close to $1$. This repeated tunneling flips the fermion parity on the quantum wire, with $i \gamma_L \gamma_R \rightarrow -i \gamma_L \gamma_R$. This can be shown to be analogous to double braiding a wire MZM along with a fictitious Majorana mode on the quantum dot and is equally robust. The periodic sign reversal of the parity operator suppresses precisely the hybridization between the Majoranas on the two ends of the wire, as desired, although there are thereotical limits to the efficacy of such a protocol. 

In this work we examine three new protocols to improve on the above proposal, discussing how they achieve both the suppression of MZM hybridization and an enhancement of the coherence of MZM qubits, how these enhancements can be measured in various experiments, and finally demonstrate how such protocols can be used to distinguish between trivial ABS bound states and MZMs. The protocols we study are: (i) The original protocol proposed in Ref.~\onlinecite{agarwalmajoranadraiding} which we name the {\it pairity-flip} protocol (ii) An enhanced protocol which we call the \textit{PH-symmetric} protocol.  This protocol can reduce the hybridization energy without being limited by the lower bound of the pairity-flip protocol which is derived from an intrinsic difference in the matrix elements for conventional and Andreev tunneling between the quantum dot and wire. It reduces the static hybridization energy significantly and can be shown to result in zero bias peaks (ZBPs) that are more pronounced and closer to zero bias voltages. (iii) The \textit{left-right-symmetric} protocol which involves coupling each end of the wire to a driven quantum dot and can be used suppress both the hybridization between majoranas, but also quasiparticle poisoning, and
(iv) a {\it composite} protocol which combines both the advantages of both the particle-hole symmetric and left-right symmetric protocols.
We test these protocols numerically with and without noise and develop the theoretical framework for studying tunneling conductance in the Floquet setting. 

All of the above protocols can also be viewed as a spin echo protocol~\cite{slichter2013principles} which suppresses noise at frequencies below the inverse timescale of successive Landau-Zener transitions, that is, the rate at which the protocols are implemented. It is well known that electromagnetic noise with a divergent low-frequency spectrum of the form $1/f$ is endemic in heterostructures realizing MZMs, and potentially limits the coherence times of putative qubits made out of weakly hybridized MZMs~\cite{knapp2018dephasing,mishmashdephasing}. Such noise can alter the hybridization energy of the Majorana qubit in time by coupling to the parity operator $-i \gamma_L \gamma_R$, and lead to dephasing. By rapidly, and periodically switching the sign of the parity operator, the spectrum of the $1/f$ noise is effectively suppressed at frequencies below the frequency at which the parity flips. We simulate numerically the coherence of qubits in nanowires hosting MZM, with experimentally relevant parameters describing the nanowire and chemical potential noise~\cite{moore2018quantized,mishmashdephasing}, and find order of magnitude or more improvement in the coherence when the protocols are implemented. 

\begin{figure}
\begin{centering}
\includegraphics[width=2.5in]{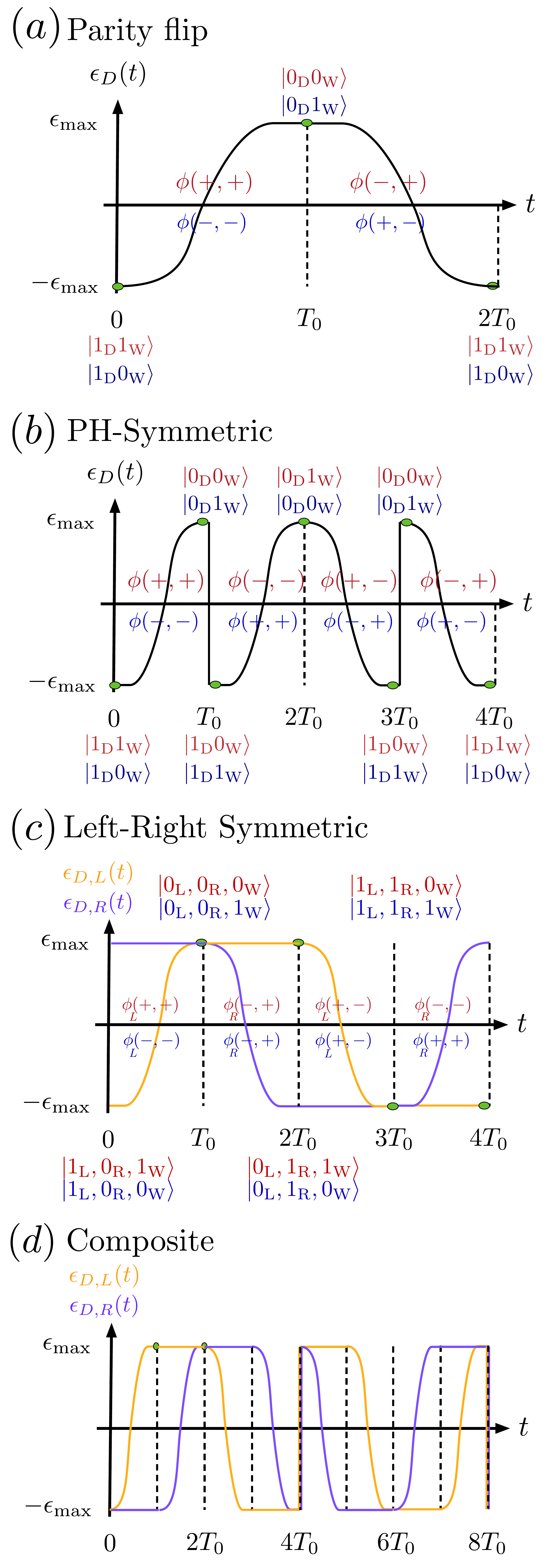}
\caption{The various protocols considered. The plots show the potential $\epsilon_D (t), \epsilon_{D,L} (t), \epsilon_{D,R} (t)$ on the quantum dot being driven [singular dot (black line) on the left for (a-b), and left (yellow) and right (purple) dots for (c-d)]. The green circles indicate the time at which the wavefunction is specified for the odd (red) and even (blue) parity states (defined by the occupation of the MZMs at $t = 0$). In the PH-Symmetry and Composite protocols, sudden parity flips on the quantum dots are performed. These parity flips must involve a fermion from another external dot which we do not describe in any more detail; here we illustrate these as sudden changes in the dot potential from $\epsilon_{\text{max}} to -\epsilon_{\text{max}}$ or vice versa. The dynamical phases $\phi_{L (R)} (\pm, \pm)$ incurred during time evolution of the two parity states are also shown, except for the Composite protocol, where it is omitted for clarity.}
\label{fig:protocols}
\end{centering}
\end{figure}

Furthermore, the unitary effecting parity switches is local (arising from the tunneling operator between the quantum dot and the wire).  If we assume the dot is coupled to the left side of the wire, it flips the sign of the Majorana operators as $\gamma_L \rightarrow - \gamma_L$, $\gamma_R \rightarrow \gamma_R$, in the process of flipping the parity. Thus, quasiparticle poisoning, which occurs when a quasiparticle from the bulk tunnels into the Majorana mode, is also suppressed at the left edge of the wire because terms involving $\gamma_L$ and a bulk fermion operator are naturally suppressed. To fully suppress quasiparticle poising, it is important to implement a protocol that also modulates the sign of $\gamma_R$ periodically in time, and out of phase with the modulation of $\gamma_L$. The \textit{left-right symmetric protocol} involving quantum dots coupled to each end of the quantum wire precisely achieves this protection against quasiparticle poisoning on both ends of the wire. The protocols are modeled by varying the potential on the quantum dot(s) in time as depicted in Fig.~\ref{fig:protocols}.

\begin{figure}
\begin{centering}
\includegraphics[width=3.1in]{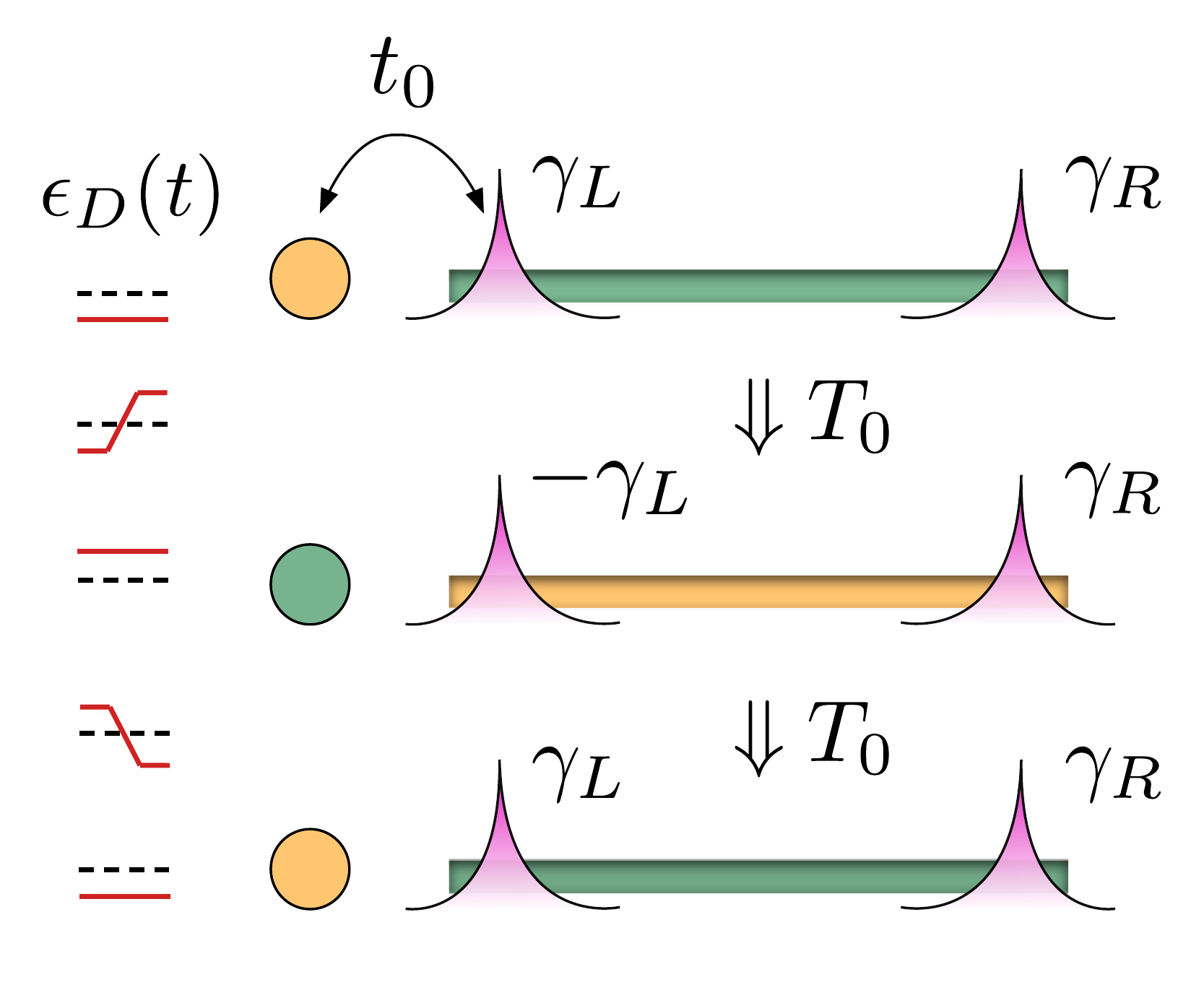}
\caption{The setup: a quantum dot (or two in case of the more elaborate protocols) is coupled to the quantum wire on one side. The potential on the quantum dot is driven periodically to induce Landau-Zener transitions between the quantum dot and the wire which flip the fermion parity on the wire, and in particular, the sign of the MZM on the left edge. }
\label{fig:exp}
\end{centering}
\end{figure}

In the above, we discussed briefly how the protocols proposed can enhance Majorana qubits, by the reduction of the hybridization between left and right Majoranas on the quantum wire, and the suppression of low-frequency dephasing and tunneling noise. A further important goal is to be able to experimentally distinguish these MZMs from trivial subgap states such as ABSs. Indeed, many theoretical studies have since shown that highly stable ABSs can be realized in quantum wires which persist for a wide range of applied magnetic fields, and at energies close to zero, making them hard to distinguish in conventional tunneling experiments from MZMs. 

Remarkably the proposed protocols can  be used as a smoking gun experiment to distinguish MZMs from ABSs and other trivial subgap states. This is due to the non-locality of the Majorana qubit---the proposed local coupling between the quantum dot and the edge of the wire hardly affects the Majorana mode energy since the qubit information is stored non-locally on both sides of the wire.  By contrast, an ABS is a regular fermionic state whose energy is accidentally close to zero.  Therefore, it is much more susceptible to local perturbations and is likely to move away from zero energy when the protocol is implemented via a time-dependent coupling to the quantum dot.  Moreover, it is important that both fermion parity states accrue the same dynamical phase over the course of periodic time evolution---it is the phase difference that determines the renormalized bound state energy. In the case of ABSs, this dynamical phase difference can be large due to the sensitivity to local perturbations. In summary, we do not expect the protocols to be effective in the case of ABSs and numerical simulations confirm this---we find that the protocols when implemented on a wire hosting ABSs in fact reduce the coherence of the putative ABS qubit instead of improving it, unlike the case when the wire hosts MZMs. 


This paper is organized as follows. In Sec.~\ref{sec:dynprotocols}, we describe the protocols proposed and present an analysis of how they reduce the hybridization energy. In Sec.~\ref{sec:Hs}, we describe the Hamiltonians for the two systems considered---that of a theoretical one-dimensional Kitaev wire composed of spinless electrons, and the more realistic spinful spin-orbit coupled semiconducting wire that describes quantum wires in experiments. In Sec.~\ref{sec:hybenergy}, we provide numerical evidence of the reduction in the hybridization energy by several orders of magnitude in the two systems we study. In Sec.~\ref{sec:conductance}, we study numerically the tunneling conductance in the presence of these dynamical protocols and the shifting of the ZBPs closer to zero voltage bias. We note that the peak height itself, whose quantization is close to $2 e^2/h$ is often experimentally used to provide evidence for the presence of Majoranas, turns out to be a poor indicator of the efficacy of the protocol and the reduced hybridization energy.  This is because the conductance quantization is fairly good even for imperfect Majoranas and this leaves little room for the protocol to make any improvement. (In App.~\ref{sec:computecond} we provide the theoretical background for computing the conductance in the Floquet setting in a numerically tractable manner.) In Sec.~\ref{sec:qubitcoherence}, we turn to measurements of the qubit coherence times with and without the protocols in the presence of $1/f$ dephasing noise. The Majorana qubit shows at least an order of magnitude improvement in the coherence time while the coherence time of qubits composed of accidental ABSs degrade and exhibit no regular oscillations. This validates our suggestion that such dynamical protocols can be used to discriminate between MZMs and ABSs. We conclude with a discussion of similar protocols in tetron qubits, and an outlook to other potential uses for such protocols in Sec.~\ref{sec:conclusions}. 

\section{Dynamical protocols}
\label{sec:dynprotocols}

We now describe the series of dynamical protocols which are designed to lower hybridization between MZMs and reduce the impact of low-frequency noise on the putative Majorana qubit. We provide here an intuition for how the protocols are designed, discuss the dynamical constraints for their implementation, and provide an analytical estimate for the reduction in the hybridization energy. 

First, we focus on the description of the protocol proposed in Ref.~\cite{agarwalmajoranadraiding} which will serve as the basis for more effective protocols we propose below. This protocol was inspired by the observation that braiding Majoranas $\gamma_i, \gamma_j$ around one another twice results in the transformation $\gamma_{i(j)} \rightarrow - \gamma_{i(j)}$, that is both Majoranas acquire a robust minus sign mandated by topological considerations. This minus sign can be used as a means for performing dynamical decoupling---the basic idea is that terms in the effective Hamiltonian, composed of these low energy Majorana modes, which involve an odd number of doubly braided Majoranas, undergo repeated oscillation of sign in time. As a result, if this double braiding is performed sufficiently rapidly (in a sense to be clarified below), then the effective Hamiltonian describing the time-evolution of the system at longer time scales, is absent of such terms. It has been proposed recently by one of the authors that such dynamical decoupling can be implemented even in many-body systems~\cite{agarwal2020polyfractal}. 

More specifically, the  MZMs $\gamma_L, \gamma_R$ at the ends of a topological quantum wire are not directly coupled to each other but are coupled through other degrees of freedom in the wire.  This creates hybridization which can be effectively represented by the term $-i \epsilon_0 \gamma_L \gamma_R$, where $\epsilon_0$ is the hybridization energy. To suppress such a term, one of $\gamma_{L (R)}$ needs to be part of a periodic double braiding with an additional Majorana mode such that it changes sign periodically causing the effective left/right coupling to average to zero over time. 

Unfortunately, performing such double braids is challenging in present experimental setups as moving Majoranas adiabatically in itself represents an enormous challenge. In order to circumvent this issue, Ref.~\cite{agarwalmajoranadraiding} proposed the use of quantum dots which can supply a fermionic mode which in turn can be interpreted as a pair of Majorana modes, albeit strongly coupled ones. (This coupling represents the energy of an electron on the quantum dot.) Nevertheless, Ref.~\cite{agarwalmajoranadraiding} showed that double braiding between Majorana modes on the quantum dot with the wire MZM can be effectively achieved with the repeated tunneling of an electron back and forth from the quantum dot. This repeated tunneling effects parity flips $i \gamma_L \gamma_R \rightarrow -i \gamma_L \gamma_R$ which can be used to suppress the hybridization between MZMs on the quantum wire. It was shown that this process is as robust as physically double braiding Majoranas whose robustness is guaranteed by topological considerations. For better readability, we revisit these arguments below. 

\subsection{Landau-Zener transitions and effect on hybridization energy}

If the potential on the quantum dot varies between values inside the bulk gap on the wire, it is unlikely for tunneling to occur into bulk modes. As a result, we can simplify our description of this system to that of tunneling between two fermionic modes---one on the quantum dot, and one comprising the MZMs on the quantum wire. Furthermore, given that the entire system conserves net fermion parity at all times, we can describe this system by two $2$-by-$2$ matrices corresponding to net (quantum dot and wire combined) even and net odd fermion parity sectors as below---
\begin{align}
H^{\pm} &= \frac{1}{2} \epsilon_D (t) +  \begin{pmatrix} \mp \epsilon_0 - \frac{1}{2} \epsilon_D (t) & \frac{t_0}{\sqrt{\xi}} (1 \pm i \beta) \\ \frac{t_0}{\sqrt{ \xi}} ( 1 \mp i \beta^* ) & \pm \epsilon_0 + \frac{1}{2} \epsilon_D (t)  \end{pmatrix}
\end{align}

where $H^+$ acts on the basis of even parity states $\ket{1_D 1_w} , \ket{0_D 0_w}$, while $H^-$ acts on the basis of odd parity states $\ket{0_D 1_w} , \ket{1_D, 0_w}$. Here $\epsilon_0$ is the hybridization energy of MZMs and $\epsilon_D (t)$ is the potential on the quantum dot that is used to drive this system---these comprise the diagonal contributions which account for the energy of the fixed parity states noted above. The off-diagonal term is proportional to $t_0$ - the amplitude of local tunneling matrix element.  The effective tunneling matrix element is suppressed by a factor of $1/\sqrt{\xi}$, where $\xi$ is the localization length of the edge Majoranas in the quantum wire. The factors of $(1 \pm i \beta)$ can be explained as follows. 

The tunnel coupling to the quantum dot occurs through the adjacent fermionic operator on the quantum wire, which we denote by $c^\dagger_L$. This operator can be decomposed into eigenoperators---among these, the only ones relevant energetically are $\gamma_L, \gamma_R$, the MZMs on the quantum wire. We thus expect 

\be
c^\dagger_1 \rightarrow \frac{1}{\sqrt{\xi}} (\gamma_L + \beta \gamma_R ) =  c_0 ( 1 + i \beta) - c_0^\dagger ( 1 - i \beta) 
\ee

where $\abs{\beta} \sim e^{- L/\xi}$ is exponentially small in the wire length $L$ as expected from locality, and we have rewritten $\gamma_L = c_0 + c^\dagger_0, \gamma_R = -i (c_0 - c^\dagger_0) $ in terms of a usual fermion operator made from the MZMs. It is important to note that $\beta$ is in general complex, and thus $\abs{1 + i \beta} \neq \abs{1 - i \beta}$. This implies that the resultant effective tunneling amplitude in the odd and even parity sectors is not identical, and will ultimately limit the efficacy of the protocol as we discuss below. Note that this difference in tunneling amplitudes physically arises from the fact that while it represents conventional tunneling in the odd parity sector (electron hopping back and forth), in the even parity sector it can be understood as Andreev tunneling where an electron from the quantum dot and the quantum wire are absorbed into the bulk and back. 

Now, the protocol is expected to be effective in the adiabatic regime wherein the wire's parity switches periodically. In this regime, the Floquet eigenstates are expected to be closely related to the instantaneous eigenstates of $H^{\pm}$ and the effective hybridization between the Majoranas is given by the time-averaged dynamical phase difference between the ground (or excited) states of $H^+$ and $H^-$, which correspond to different parity sectors in the wire away from the Landau-Zener transitions. This follows analogously to how the hybridization energy in the static setting can be ascertained from the rate of phase accrual between ground states of different parities. In particular, we find

\begin{align}
\epsilon_n &= \frac{1}{2T_0} \abs{\int_0^{2 T_0} dt \;\left[E^+ (t) - E^- (t)\right]}, \; \; \text{where} \nonumber \\
E^{\pm} &= \frac{\epsilon_D (t)}{2} - \sqrt{ \left( \pm \epsilon_0 - \frac{\epsilon_D (t)}{2} \right)^2 + \frac{t^2_0}{\xi} \abs{1 \pm i \beta}^2 }    
\label{eq:Epm}
\end{align}
are the energies of the ground states of $H^\pm$, and $\epsilon_n$ is the renormalized hybridization energy between the MZMs. 

Note that all protocols we study satisfy the condition $\epsilon_D (t) = - \epsilon_D (t + T_0)$, where $T_0$ is half the period of the protocol. For $\beta = 0$, it is easy to see that $\epsilon_n = 0$. One can estimate $\epsilon_n$ more generally by noting that the phase difference is primarily accrued when the tunneling matrix element is comparable to the energy. This occurs for two periods of duration $\sim \frac{t_0 T_r}{ \epsilon_{\text{max}} \sqrt{\xi}} $ within the $2T_0$ period of the protocol, where $T_r < T_0$ is the ramp time associated with the Landau-Zener transition, and $\pm \epsilon_{\text{max}}$ is the maximum/minumum  potential on the quantum dot. During these ramp times, the rate of accruing phase is $\sim \frac{t_0 \abs{\beta}}{\sqrt{\xi}}$. Further estimating $\epsilon_0 \sim \Delta e^{-L/\xi} \sim \Delta \abs{\beta}$, where $\Delta$ is the bulk  gap, we find 

\be
\epsilon_n \sim \frac{t_0 \abs{\beta}}{\xi} \cdot \frac{t_0 T_r}{\epsilon_{\text{max}} T_0} \Rightarrow \frac{\epsilon_n}{\epsilon_0} \sim \frac{1}{\xi} \cdot \frac{T_r}{T_0} \cdot \frac{t^2_0}{\Delta \epsilon_{\text{max}}}
\ee

We emphasize this result is valid only when the parameter $x_{\text{LZ}} = \frac{t^2_0 T_r}{\hbar \xi \epsilon_{\text{max}}}$ that controls the adiabaticity of the Landau-Zener transitions is significantly greater than $1$. Furthermore, we require generally $t_0/\sqrt{\xi} < 2 \epsilon_{\text{max}}$, that is the maximum off-diagonal matrix element is smaller than maximum diagonal piece (to maintain the sanctity of the Landau-Zener crossing). This implies 

\be
\frac{\epsilon_n}{\epsilon_0} \sim x_{\text{LZ}} \frac{\hbar/T_0}{\Delta} \gtrsim \frac{\epsilon_0}{\Delta}. 
\label{eq:majsupp}
\ee

In summary, the best suppression is achieved for $x_{\text{LZ}} \gtrsim 1$, and \emph{largest} feasible drive period $T_0 \sim 1/\epsilon_0$ determined by the original Majorana hybridization. Thus, the protocol works better when one already has reasonably well localized Majorana modes. For the present experimental state of quantum nanowries harboring MZMs, this can represent a suppression of a few orders of magnitude. 

[Note that in the above we used that the minimum driving frequency to avoid rapid heating~\cite{abanin2016theory,MoriSaitoKuwahara,agarwal2020polyfractal} dictated by the hybridization energy of MZMs, but in the context of Majoranas specifically, it may be possible to drive slower without heating the system because the Majoranas represent an energetically isolated finite size system (with the Hilbert space dimension of a qubit) where heating can be less prominent even when driving at lower frequencies~\cite{khodjastehlidarviolaPRL1}.]

\subsection{Improved protocols}

We now propose improvements to the protocol proposed in Ref.~\cite{agarwalmajoranadraiding}. First, let us partition the dynamical phase picked up by the MZMs in Eq.~(\ref{eq:Epm}) into two parts---an integral from $0$ to $T_0$, and an integral from $T_0$ to $2T_0$. Utilizing the condition that $\epsilon_D (t + T_0) = - \epsilon_D (t)$, the phase difference between even and odd parity states results in the effective hybridization energy given by

\begin{align}
    \epsilon_n &= \abs{ \left[ \phi(+,+)+\phi(-,+) \right] - \left[ \phi(-,-) + \phi(+,-) \right]}, \nonumber \\
    \phi(a,b) &= \frac{-1}{2T_0} \int_0^{T_0} dt \sqrt{\left(a \epsilon_0 - \frac{\epsilon_D (t)}{2} \right)^2 + \frac{t^2_0}{\xi} \abs{1 + ib \beta}^2},  
\end{align}

where $a, b = \pm 1$. The above expression highlights the role of the tunneling matrix element $t_0 \abs{1 \pm i \beta}$ which differs between conventional tunneling and Andreev tunneling in the even and odd parity sectors, and in general prevents the phase accrued in the even parity sector, given by $\phi(+, +) + \phi (- , + )$ from agreeing with that accrued in the odd parity sector, given by $\phi(-,-)$ + $\phi(+,-)$. This limits the reduction of the effective hybridization energy to a factor of $\beta \sim e^{-L/\xi}$ as noted above. 

We propose below three protocols that aim to rectify this issue and lead to further suppression of the hybridization energy. 


\subsubsection{PH-symmetric protocol}

In the `PH-symmetric' protocool, we explicitly rectify the above issue by changing the parity (fermion number) on the quantum dot twice during the time period of the Floquet driving. This parity flipping may be achieved, say, using another quantum dot; for the sake of simplicity, we do not model this additional quantum dot. This parity flip on the quantum dot forces both conventional and Andreev tunneling to take place in both even and odd parity sectors, symmetrizing the net phase accrued in both cases. 

Taking Fig.~\ref{fig:protocols}b for reference, following the path taken by the even and odd parity states over one Floquet period, we can confirm that the phase accrued by the system when starting in the odd parity sector is given by $\phi(-,-) + \phi(+,+) + \phi(-,+) + \phi(+,-)$ which matches exactly the phase accrued by the even parity state. 

In this case, we anticipate that the effective hybridization energy is largely limited by the non-adiabaticity of the Landau Zener transitions.  


\subsubsection{Left-Right Symmetric protocol}

One of the key goals of performing such protocols is to not just reduce the static hybridization between Majoranas but also to reduce the deleterious impact of low-frequency noise on the system. In general, the above protocols lead to the periodic transformation of MZMs according to the rule $\gamma_L \rightarrow - \gamma_L, \gamma_R \rightarrow \gamma_R, -i \gamma_L \gamma_R \rightarrow + i \gamma_L \gamma_R$. (Numerical evidence of the sign flipping only on $\gamma_L$ was provided in Ref.~\cite{agarwalmajoranadraiding}.) Here we note that we've noted the locality of the Landau Zener transitions in asserting that it is the Majorana closer to the quantum dot (in this case, $\gamma_L$) that assumes the minus sign from parity flipping operations. This implies that all low-frequency dephasing noise, which couples to $i \gamma_L \gamma_R$ (of which the static hybrdization can be viewed as a special limit) is suppressed but also any \emph{quasiparticle poisoning} which may occur via a coupling of a fermionic bath operator with $\gamma_L$. However, quasiparticle poisoning via $\gamma_R$ remains uncontrolled. 

It may therefore be desirable to find a protocol that suppresses quasiparticle poisoning on both ends of the wire. We propose such a protocol and term it the `left-right symmetric' protocol. In particular, two quantum dots, one on each end of the quantum wire are employed in this case. Of course, we must now account for the differences in the tunneling amplitudes on the left and right ends of the wire. We will thus add a subscript $L (R)$ to the phase accrued $\phi_{L (R)} (\pm, \pm)$ defined above to denote this difference. 

The protocol is pictorially depicted in Fig.~\ref{fig:protocols}c. We see that the phase picked up in the odd parity sector is given by $\phi_L (-,-) + \phi_R (+, -) + \phi_L (-,+) + \phi_R (+, +)$, while that in the even parity sector is given by $\phi_L (+,+) + \phi_R (-,+) + \phi_L (+,-) + \phi_R (-,-)$. We note that the phases accrued will in general be different from one another (and result in a hybrdization energy reduction of similar magnitude to the protocol proposed in Ref.~\cite{agarwalmajoranadraiding}) unless the tunneling amplitudes from dots at the left and right ends of the system are carefully tuned to match each other. In the case of perfect matching, the protocol should lead to the reduction in hybridization energy of a similar order of magnitude compared to the PH-symmetric protocol.   


\subsubsection{Composite protocol}

We finally propose a `composite' protocol which combines the advantages of the left-right symmetric protocol in eliminating low-frequency quasiparticle-poisoning noise and dephasing noise from both ends of the wire, and the PH-symmetric protocol in suppressing the static hybridization energy between MZMs without requiring fine-tuning. On the flip side, the protocol involves a quantum dot on either end of the wire, with possibly another quantum dot on each side for flipping the parity of each of these two quantum dots on demand. The full protocol is depicted in Fig.~\ref{fig:protocols}d. It is evident that the phase picked up by the even and odd parity modes over the course of a Floquet period is identical regardless of differences in the coupling strengths on either end of the wire. The downside of the protocol is of course it is much more complicated to implement and involves longer time periods which thus limits the ability to suppress noise at moderate frequencies. Note that we do not simulate this protocol numerically but we have listed the principle behind deriving it for completeness. 


\section{Hamiltonians studied}
\label{sec:Hs}
To numerically verify our ideas, we consider two Hamiltonians to examine the effect of proposed dynamical protocols. First, for simplicity, we consider the Kitaev chain~\cite{kitaev2001unpaired} which describes a spinless $p$-wave superconductor given by the Hamiltonian
\begin{equation}
H_K = -\mu\sum^L_{n=1}c^\dagger_nc_n-\frac{1}{2}\sum^{L-1}_{n=1}\left(tc^\dagger_nc_{n+1}+\Delta c_nc_{n+1}+h.c.\right), 
\end{equation}
where $L$ is the length of the chain, $\mu$ is the chemical potential, $t$ is the nearest neighbour hopping amplitude, and $\Delta$ is the superconducting pairing amplitude. This system is well known to exhibit MZMs in the topological phase characterized by $\abs{\mu} < t$. For $\mu = 0$, the MZMs are perfectly localized on one site and isolated from one another. The MZMs develop width and exhibit increasing overlap as one approaches the transition at $\mu \rightarrow t$ from below, or $\mu \rightarrow -t$ from above.

We also study the more experimentally relevant model of a semiconducting nanowire with strong spin-orbit coupling proximited by a conventional $s$-wave pairing superconductor under an external magnetic field. This Hamiltonian can loosely be understood as a doubled copy of the Kitaev chain, with an external magnetic field that elements one set of Majorana edge modes in order to render the system in the Kitaev phase. The discretized version of this Hamiltonian, which we employ to perform numerical simulations, is given by

\begin{widetext}
\begin{equation}
\begin{aligned}
H_\text{NW} = -&\frac{t}{2}\sum^{L-1}_{n\sigma}\left(c^\dagger_{n\sigma}c_{n+1,\sigma}+h.c.\right)+\sum^L_{n\sigma} \left(t-\mu + V_n \right) c^\dagger_{n\sigma}c_{n\sigma} - B_z \sum^L_{n\sigma\sigma'}c^\dagger_{n\sigma}(\sigma^z)_{\sigma\sigma'}c_{n\sigma'}\\
+&\frac{\alpha}{2}\sum^{L-1}_{n\sigma\sigma'}\left(c^\dagger_{n\sigma}(i\sigma^y)_{\sigma\sigma'}c_{n+1,\sigma'}+h.c. \right)+\sum^L_{n}\Delta_n\left(c_{n\uparrow}c_{n\downarrow}+c^\dagger_{n\downarrow}c^\dagger_{n\uparrow}\right)
\end{aligned}
\end{equation}
\end{widetext}

where $t \equiv \hbar^2/ma^2$ is the effective nearest neighbour hopping ($a$ is the lattice spacing of a sample and $m$ is the effective electron mass), $\mu$ is the chemical potential, $V_n$ is the site dependent potential, $B_z$ represents the Zeeman energy due to an external magnetic field in the $z-$direction, $\alpha \equiv \alpha_R/a$ is the spin-orbit strength ($\alpha_R$ being the amplitude of Rashba spin-orbit coupling in the continuum model), and $\Delta_n$ is the site dependent pairing amplitude. When simulating the nanowire in a phase where MZMs are harbored, we set $V_n = 0$ and $\Delta_n = \Delta$.  A transition from a trivial to a topological phase is controlled by the Zeeman energy where the criterion for a topological phase is given by $\abs{B_z}>\sqrt{\mu^2+\Delta^2}$.

The purpose of the site dependence for $V_n$ and $\Delta_n$ is to introduce a suppression of superconducting correlations near the end of a wire and employ a potential to trap low energy ABSs. Following Ref.~\cite{mishmashdephasing}, we use the following forms for these terms--- 
\begin{subequations}
\begin{equation}
V_n=\frac{V_0}{2}\left[-\tanh(\frac{na-x_0}{l_V})+1\right],
\end{equation}
\begin{equation}
\Delta_n=\frac{\Delta_0}{2}\left[\tanh(\frac{na-x_0}{l_\Delta})+1\right].
\end{equation}
\end{subequations}
More details of the precise parameters will be specified in Sec.~\ref{sec:hybenergy}.

The QD Hamiltonian and coupling to the wire is given by
\begin{equation}
H_d = \epsilon_d(t)c^\dagger_dc_d-t_0\left(c^\dagger_dc_1+h.c.\right)
\end{equation}
for the Kitaev chain. Here $\epsilon_d(t)$ is the local potential on the QD (see Fig. \ref{fig:Local_potentials}) and $t_0$ is the coupling between the QD and the left edge of the Kitaev chain. In the case of the left-right symmetric protocol, we simply assume the same parameters for another quantum dot coupled to the right edge of the Kitaev chain. In the case of the semiconducting nanowire, we also have a magnetic field on the quantum dot
\begin{align}
H_d &= \left( \epsilon_d (t) + V_z \right) \sum_{\sigma} c^\dagger_{d \sigma} c_{d \sigma} - V_z \left(c^\dagger_{d \uparrow} c_{d \uparrow} - c^\dagger_{d \downarrow} c_{d \downarrow} \right) \nonumber \\
& - t_0 \sum_\sigma \left( c^\dagger_{d \sigma} c_{1 \sigma} + \text{h.c.} \right)
\end{align}
where we assume the same magnetic field applies to the QD as the wire. Note that the Zeeman coupling may be different on the quantum dot and the quantum wire depending on their effective g-factors. Although we have not accounted for this difference, we note that the effect of changing this Zeeman coupling on the quantum dot is minimal---it is merely important to correct for the shift by adding an extra spin-independent potential $V_z$ on the QD to make sure that the desired spin mode lies within the bulk gap and crosses the MZM energies. 

\section{Reduced Hybridization energy}
\label{sec:hybenergy}
In this section, we present numerical verification of the efficacy of the above proposed protocols to reduce the hybridization energy of the MZMs. We consider both the Kitaev chain and the more realistic spin-orbit coupled nanowire supporting MZMs. We generally see a much greater efficacy of the PH-symmetric and Left-Right Symmetric protocols in reducing the hybridization between MZMs as compared to the simple Parity Flip protocol proposed in Ref.~\cite{agarwalmajoranadraiding}. 

\subsection{Kitaev chain}
\begin{figure}[htp]
\includegraphics[width=3.2in]{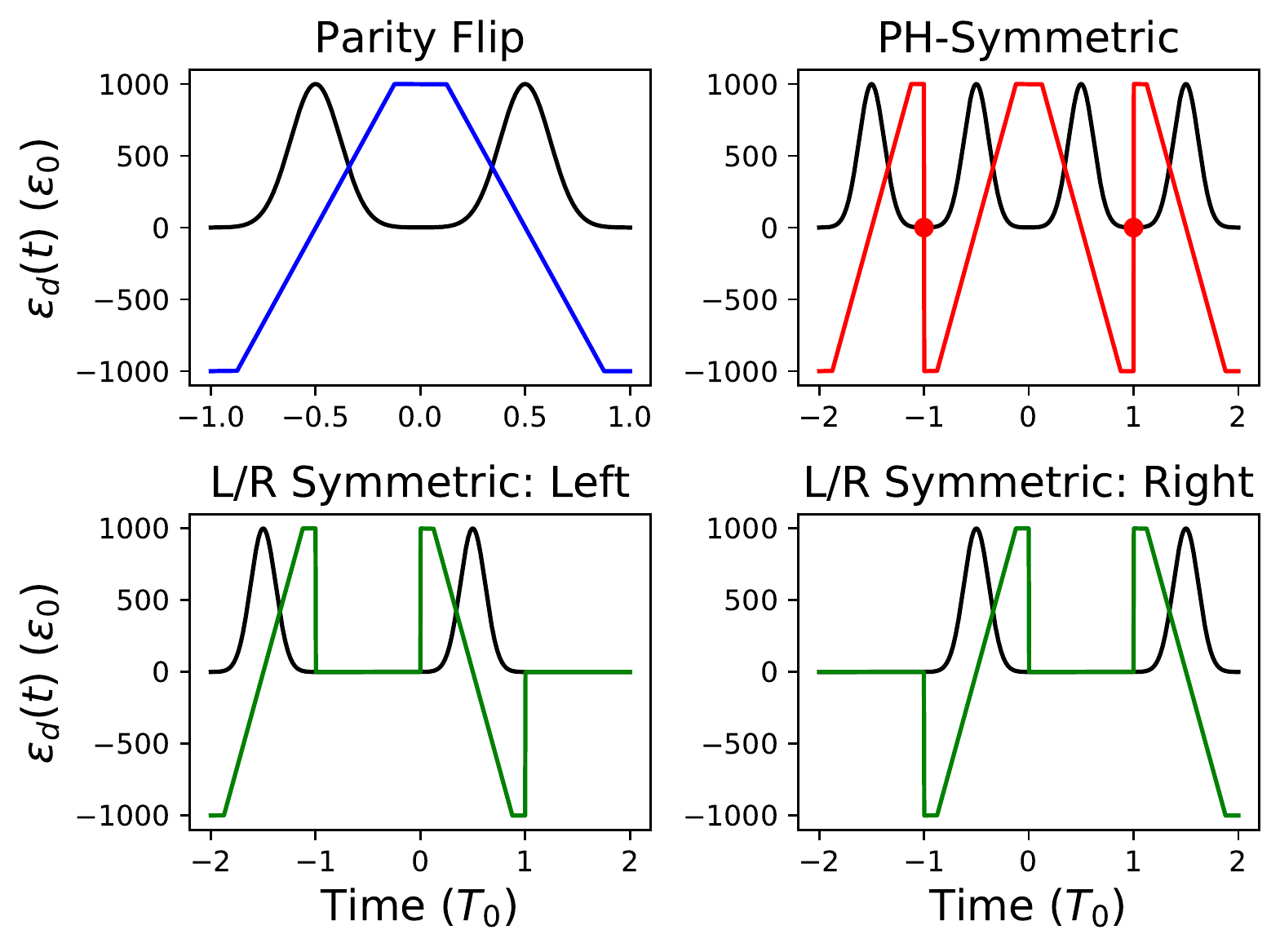}
\caption{Local potential (color) on the QD and tunnel coupling with wire (black) for different protocols for a chain of length $L=50$,  $\mu=1.72$, $t=\Delta=2$ in the Kitaev phase. The driving period $T_0=3/4\epsilon_0$, the ramp time $T_r=3T_0/4$, the time-dependent QD-wire coupling corresponds to superposed Gaussians of width $\sigma=0.16 T_0$, and the potential swings on the QD from  $\epsilon_{\text{max}}=1000\epsilon_0$ to $-\epsilon_{\text{max}}$. For $\mu=1.8$, we have a driving period of $T_0=3/2\epsilon_0$, a ramp time $T_r=17T_0/20$, and maximum potential variation on the quantum dot $\epsilon_{\text{max}}=120\epsilon_0$ while other parameters are the same as the $\mu=1.72$ case. For the PH-Symmetric protocols, it is at times marked by red dots where we perform the particle-hole transformation on the QD, accompanied by a rapid change in the potential on the QD. }
\label{fig:Local_potentials}
\end{figure}
In Fig.~\ref{fig:Local_potentials}. we specify the parameters on the QD local potentials and its coupling to the Kitaev chain, in units of the static hybridization energy $\epsilon_0$. The amplitude for local potentials $\epsilon_{\text{max}}$ are chosen such that the QD potential lies in between the bulk gap of the Kitaev chain, and periodically crosses the MZM energy. This ensures that electron on the QD is unable to access above-gap states, and tunneling only occurs into the MZM, if it occurs at all. We simulate the Kitaev chain with two representative values of the chemical potential for which the MZM hybridization energy is a factor of $10^2$ and $10^3$ less than the bulk gap. It is believed that the former represents the status of experimentally realized MZMs. More concretely, we simulate a chain of length $L=50$, the tunneling and superconducting order parameter of $t=\Delta=2$, and $\mu = 1.72, 1.8$ with the (former) latter value yielding MZMs with a hybridization energy of ($10^{-3}$) $10^{-2}$ times the bulk gap as determined numerically. 

We generally find that a larger $\epsilon_{\text{max}}$ close to, but smaller than the bulk gap, enables the best adiabatic Landau-Zener transitions. Also, adiabaticity is maintained best when the driving time period is large, of the order of the static hybridization energy or even higher. Additionally, we allow the coupling between the quantum dot and the wire to be time dependent, and assume it takes a Gaussian shape in time, centered around the crossing of the QD energy and MZM energy, with a width $2 \sigma$ of the order of the ramp time $T_r$. This ensures that when the quantum dot is energetically detuned from the MZM, the coupling between the dot and the wire is zero. The latter prevents any long term generation of entanglement between the quantum dot and the MZMs on the quantum wire. The precise parameters and their time-dependent profiles are illustrated in Fig.~\ref{fig:Local_potentials}. 

Numerically, we compute the Floquet unitary describing the time evolution of the coupled quantum dot and quantum wire system. Upon exact diagonalization, it is necessary to identify the correct mode that represents the Floquet MZMs using a computation of the inverse participation ratio (IPR). This is necessary because of the folding of Floquet quasi-energies into the limited frequency domain $\in (-\pi/2T_0, \pi/2 T_0]$ or $\in (-\pi/4T_0, \pi/4 T_0]$ depending on whether we consider the parity-flip protocol, or the PH/left-right symmetric protocols. This frequency domain is much smaller than the bulk excited states which are effectively static---their energy is then translated by many multiples of the Floquet drive frequency into the above frequency domain and can thus have a quasi-energy arbitrarily close to zero. This is simply a trivial consequence of the Floquet spectrum folding and so the correct MZM mode cannot be simply identified as the one with the lowest quasi-energy. We select the MZM mode by maximizing the weight of the Floquet eigenstate on the wire and this generically works well.

\begin{figure}[htp]
\includegraphics[width=3.2in]{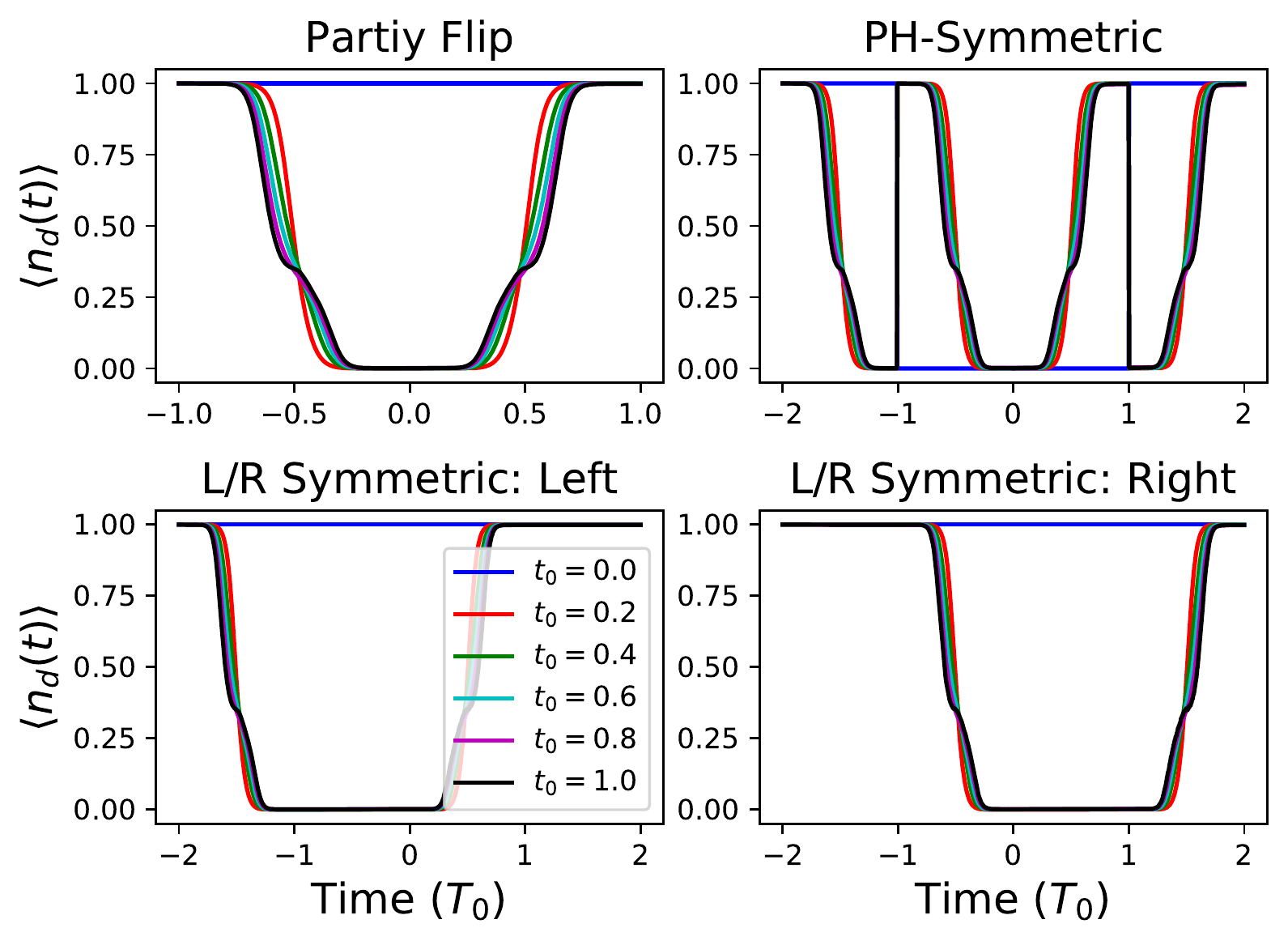}
\caption{Efficiency of LZ transition for proposed protocols at various coupling amplitudes $t_0$ in  units of $2\epsilon_\text{max}$. For all the protocols, we enter the ideal LZ regime around $t_0=0.2$. }
\label{fig:LZ_trans}
\end{figure}

In Fig.~\ref{fig:LZ_trans}, we plot the electron density on the quantum dot in time, starting from an initial state in which the quantum dot houses an electron, and averaging over all possible initial states of the quantum wire (including those where the MZM may be occupied or empty). This is computed as
\begin{align}
\label{eq: electron density}
\left\langle n_d(t)\right\rangle &= \frac{1}{Z_w} \sum_w\bra{\psi_w}\otimes\bra{1_d} c^\dagger_d(t)c_d(t)\ket{1_d}\otimes\ket{\psi_w} \nonumber \\
&= \frac{1}{2}+\frac{1}{2}\left(\abs{u_d}^2-\abs{v_d}^2 \right), 
\end{align}
where $\ket{\psi_w}$ is an arbitrary wire state and $\ket{1_d}$ is the initial wavefunction that is the occupied QD state, and $u_d$ and $v_d$ represent, respectively, the particle and hole content of quantum dot fermion operator $c_d (t)$ at time $t$ on the quantum dot. A similar expression is employed for the realistic wire setup which has analogous form for $\left\langle n_d(t)\right\rangle$ that includes spin degrees of freedom; for details, see Appendix \ref{sec:detailsontracking}.

\begin{figure}[htp]
\includegraphics[width=3.2in]{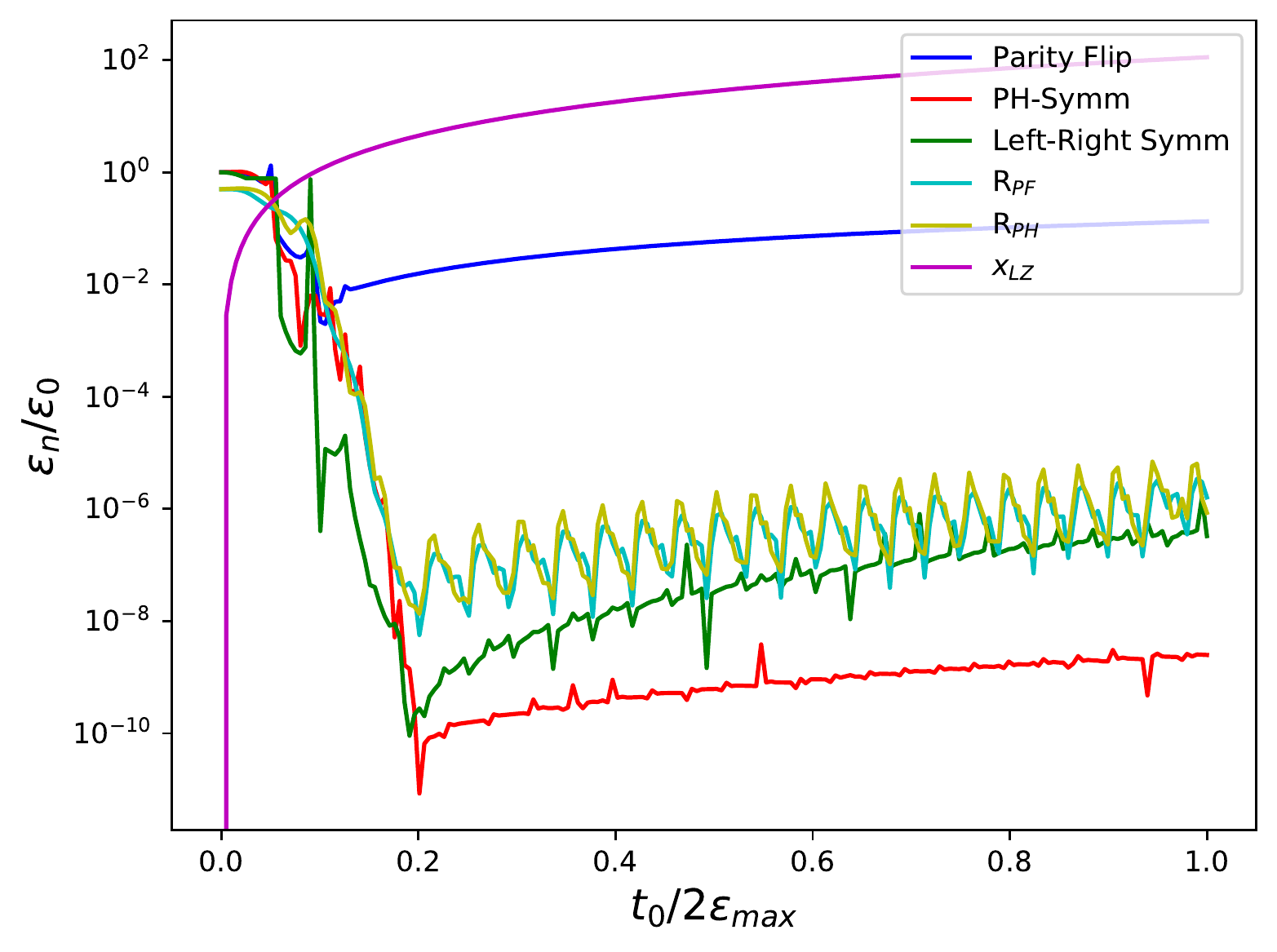}
\caption{The hybridization curve for $\mu = 1.72$. The static hybridization energy is three orders of magnitude smaller than the bulk gap in this case. The driving period $T_0 =3/4\epsilon_0 $.}
\label{fig:kitaevhyb1}
\end{figure}
\begin{figure}[htp]
\includegraphics[width=3.2in]{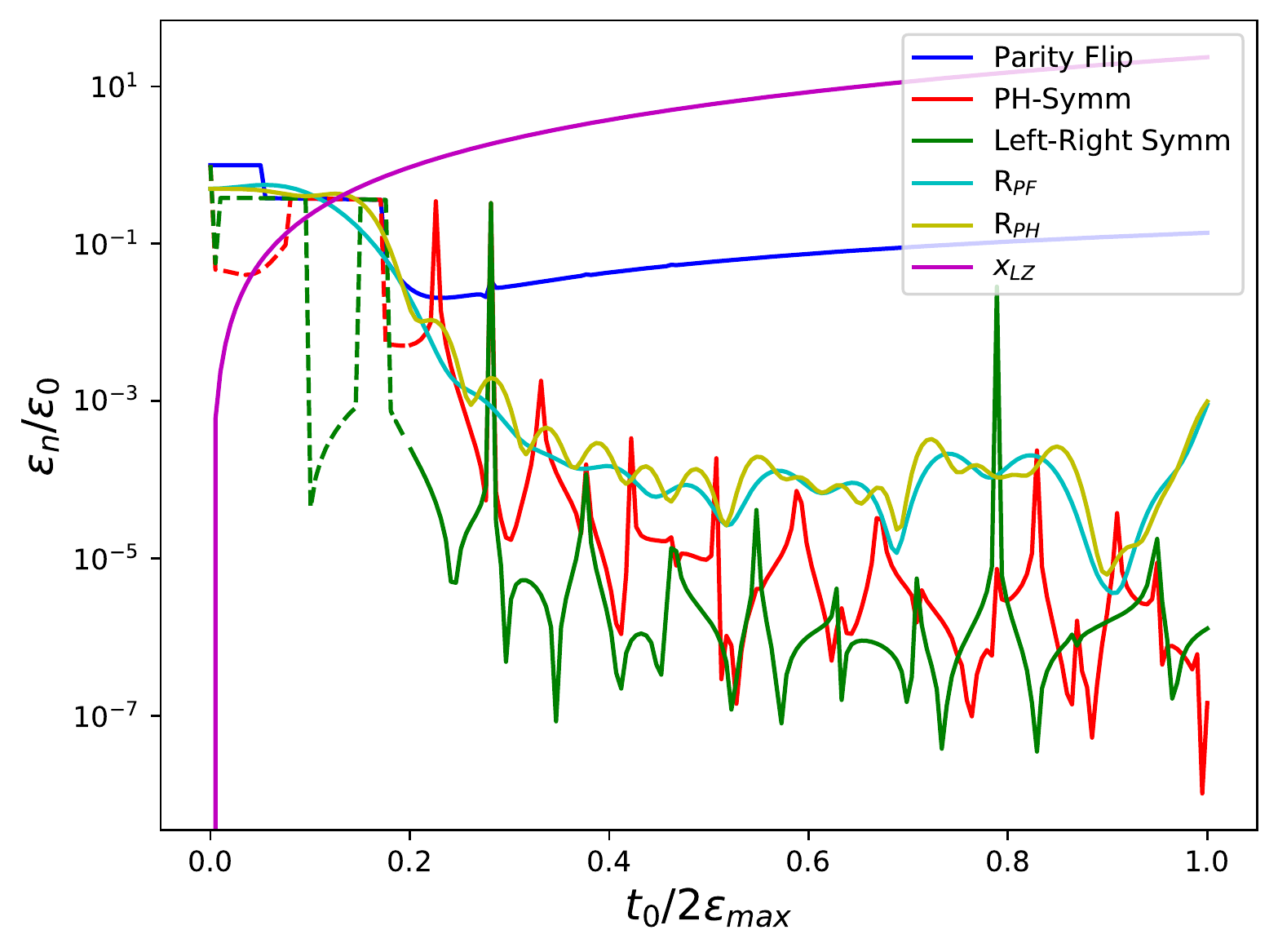}
\caption{The hybridization curve for $\mu = 1.80$. The static hybridization energy is two orders of magnitude smaller than the bulk gap in this case. The driving period $T_0 =3\epsilon_0/2 $.}
\label{fig:kitaevhyb2}
\end{figure}
\begin{figure}[htp]
\includegraphics[width=3.2in]{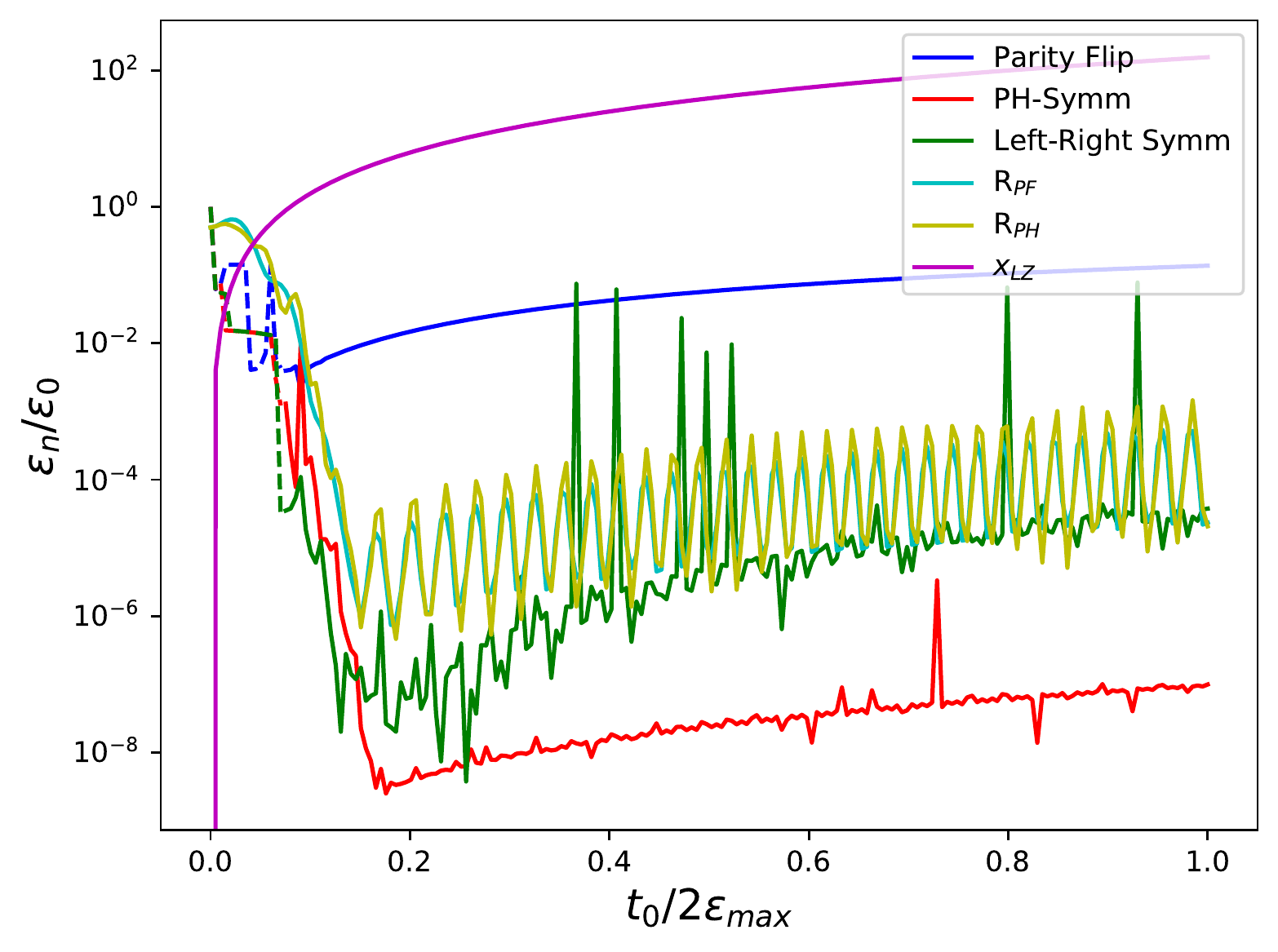}
\caption{The hybridization curve for $\mu = 1.80$ with a larger driving period $T_0 =10/\epsilon_0 $. We see a greater suppression in this case for the PH-symmetric and Left-Right Symmetric protocols in approximate agreement with the adiabaticity of the Landau-Zener transitions for Parity Flip and PH-Symmetric protocols, $R_{PF}$ and $R_{PH}$.}
\label{fig:kitaevhyb3}
\end{figure}

Figs.~\ref{fig:kitaevhyb1},~\ref{fig:kitaevhyb2}, and~\ref{fig:kitaevhyb3} show the ratio of the renormalized hybridization energy to the static hybridization energy, $\epsilon_n/\epsilon_0$, as a function of the peak coupling strength between the quantum dot and the wire, $t_0$ (in units of $2 \epsilon_{\text{max}}$) for the Parity-Flip protocol, the PH-symmetric protocol, and the Left-Right symmetric protocol. In addition, we plot $x_{\text{LZ}}$, the numerically estimated Landau-Zener parameter, and a figure of merit, $R_{PF}$ and $R_{PH}$ that captures deviations from adiabaticity in the Partiy Flip and PH-Symmetric protocols respectively. This figure of merit is computed as an average over deviations from $1$ or $0$ of the amplitude to find the electron on the quantum dot $\avg{n_d (t)}$ plotted in Fig.~\ref{fig:LZ_trans} at times chosen in the middle of successive Landau Zener transitions. That is 
\begin{equation}
R_{PF}=\frac{(1-\langle n_d(T_0) \rangle)+\langle n_d(0)\rangle}{2}
\end{equation}
and $R_{PH} = $
\begin{equation}
\frac{\langle n_d(-T_0-\delta)\rangle + \langle n_d(0) \rangle+\langle n_d(T_0 - \delta)\rangle+(1-\langle n_d(2 T_0)\rangle)}{4} 
\end{equation}
where $\pm T_0 - \delta$ are times right before the PH transformation. For ideal LZ transition, we expect $R_{PF}=R_{PH}=0$. As expected, in all cases, we find that this probability does not control the suppression of hybridization in the parity-flip protocol, but closely trails the probability of non-adiabatic transitions for the other protocols proposed. This is because in the latter case, the phase accrual is symmetrized and likely not the determining factor for the finite renormalized hybridization of the Floquet MZMs. We can also surmise that the best suppression achievable by the parity-flip protocol for $T_r \sim T_0$ is given by the ratio of the static hybridization energy to the bulk gap---for $\mu = 1.72$, the maximum suppression achieved is of the order of $10^3$, while for $\mu = 1.8$, the maximum suppression achieved is of the order of $10^2$. We see generally that the hybridization energy is much more suppressed for the PH-symmetric and Left-Right symmetric protocols. Finally, we note that there appear to be some non-analytical features in both $R_{PF}$ and $R_{PH}$ and correspondingly the suppression of hybridization in the PH-symmetric and Left-Right-symmetric protocols. We anticipate this could be due to subtle interference effects (see also App.~\ref{sec:deviationLZ} for more details) which may give rise to local dips in the adiabatic transition probability; these appear to depend highly on the precise waveform used for modulating the potential on the quantum dot.


\subsection{Semiconducting nanowire}

We now consider simulations on the spinful semiconducting wire with large spin-orbit coupling. In this case, we will use the parameters from Refs.~\cite{mishmashdephasing} which are close to the experimental realizations of MZMs in semiconducting nanowires. These parameters are, concretely, $L=200$ sites, $\Delta=0.25$, $\mu=3.8\Delta$, $t=25.4$, $\alpha=-5$, and $V_z=1.4$ where all energy are in units of meV. We also examine how the protocol works in the case of ABSs which have energies similar to that of MZMs. In this case, the parameters used are $L=200$ sites, $\Delta_0=0.25$, $\mu=3.8\Delta_0$, $V_0=3.8\Delta_0$, $t=25.4$, $\alpha=-5$, $V_z=0.7143$, $x_0=0.3$, $l_V=0.03$, $l_\Delta=0.03$, and $a=0.01$ where again, all the energies are in units of meV and all the lengths are in units of $\mu m$. The length parameters are for on-site potential and superconducting pairing amplitude introduced in \ref{sec:Hs}. Note that, in order for the electron to now tunnel into the quantum wire, it's spin must additionally align with the spin direction of the MZMs. This is automatically achieved by the external magnetic field which is applied on the quantum wire, which we will also assume to apply to the quantum dot equally. 

The simulations on the semiconducting nanowire mirror the results of the Kitaev chain. In particular, the large external magnetic field applied (which we assume to be the same on the dot as on the nanowire) effectively eliminates the spin state in the opposite direction on the quantum dot. As long as the electronic spinor on the quantum dot has finite overlap with the MZM spin configuration, Landau Zener transitions can be achieved. We find this to be the case for the experimentally relevant parameters. 

Crucially, as was suggested above, the Landau Zener transitions are not effective when it comes to ABSs. The reason is that ABSs are easily perturbed by a local perturbation. Thus, when the tunnel coupling between the quantum dot and nanowire is turned on to effect the Landau Zener transition, the ABSs drift in energy and it is hard to achieve perfectly adiabatic transitions which transfer the electron to the wire with near $1$ probability. Instead, as a consequence of imperfect tunneling, the ABSs end up hybridizing with the quantum dot. In particular, this implies that the Floquet eigenstates exhibit entanglement between the quantum dot and the quantum wire---this in general affects the hybridization energy of the ABS which we attempt to identify by maximizing the IPR on the quantum wire (although it is much less evident what should be deemed the ABS after it gets entangled with the quantum dot). As we will see, the autocorrelators of the effective qubit formed from the ABSs do not exhibit clear oscillations at a single frequency when the protocols are implemented. 

\begin{figure}[htp]
\includegraphics[width=3.2in]{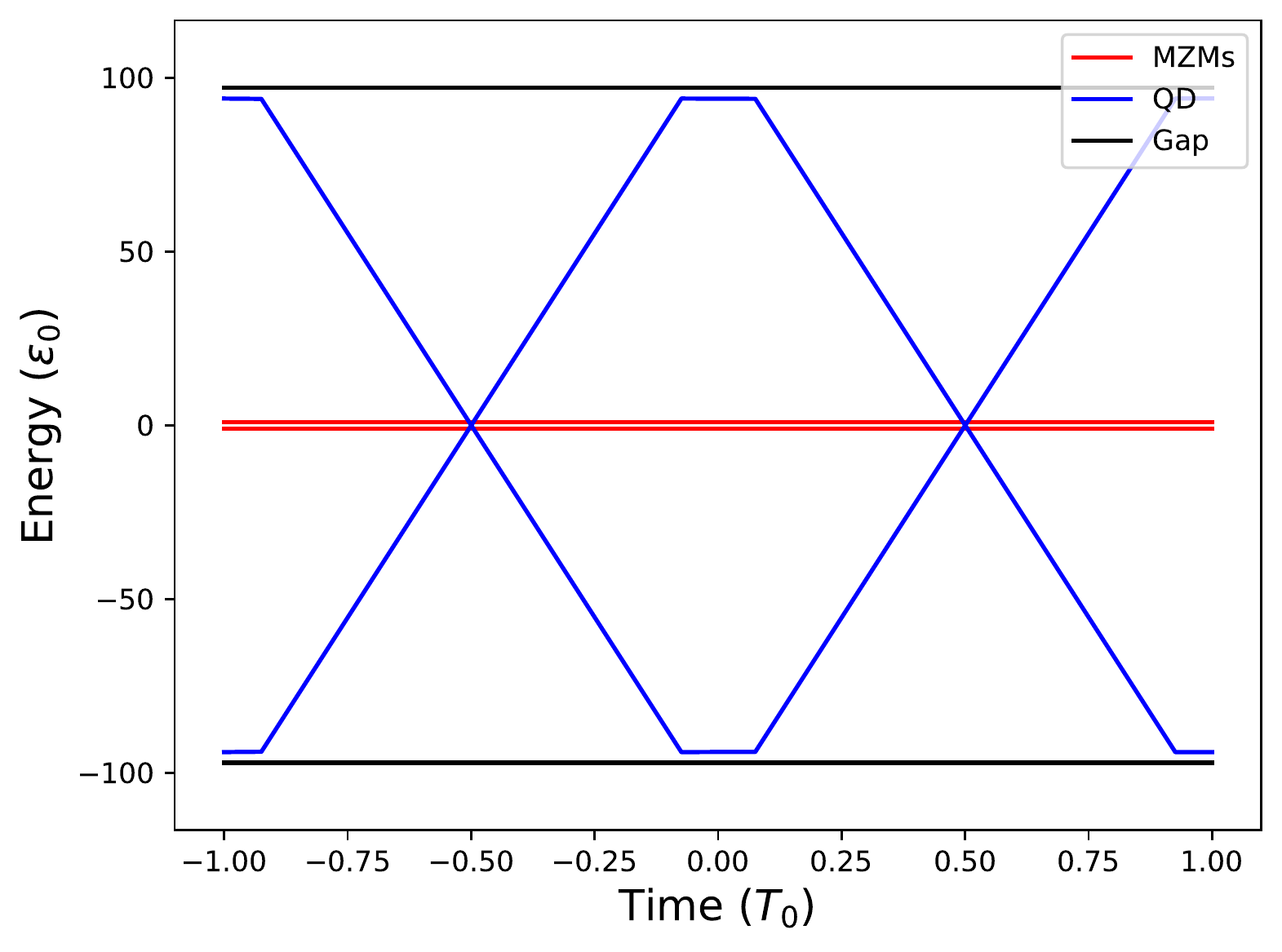}
\caption{QD, gap, and MZMs instantaneous eigenenergies. Spin down states lies within the gap while spin up states lie far outside the gap.}
\label{fig:nanopotentials}
\end{figure}

\begin{figure}[htp]
\includegraphics[width=3.0in]{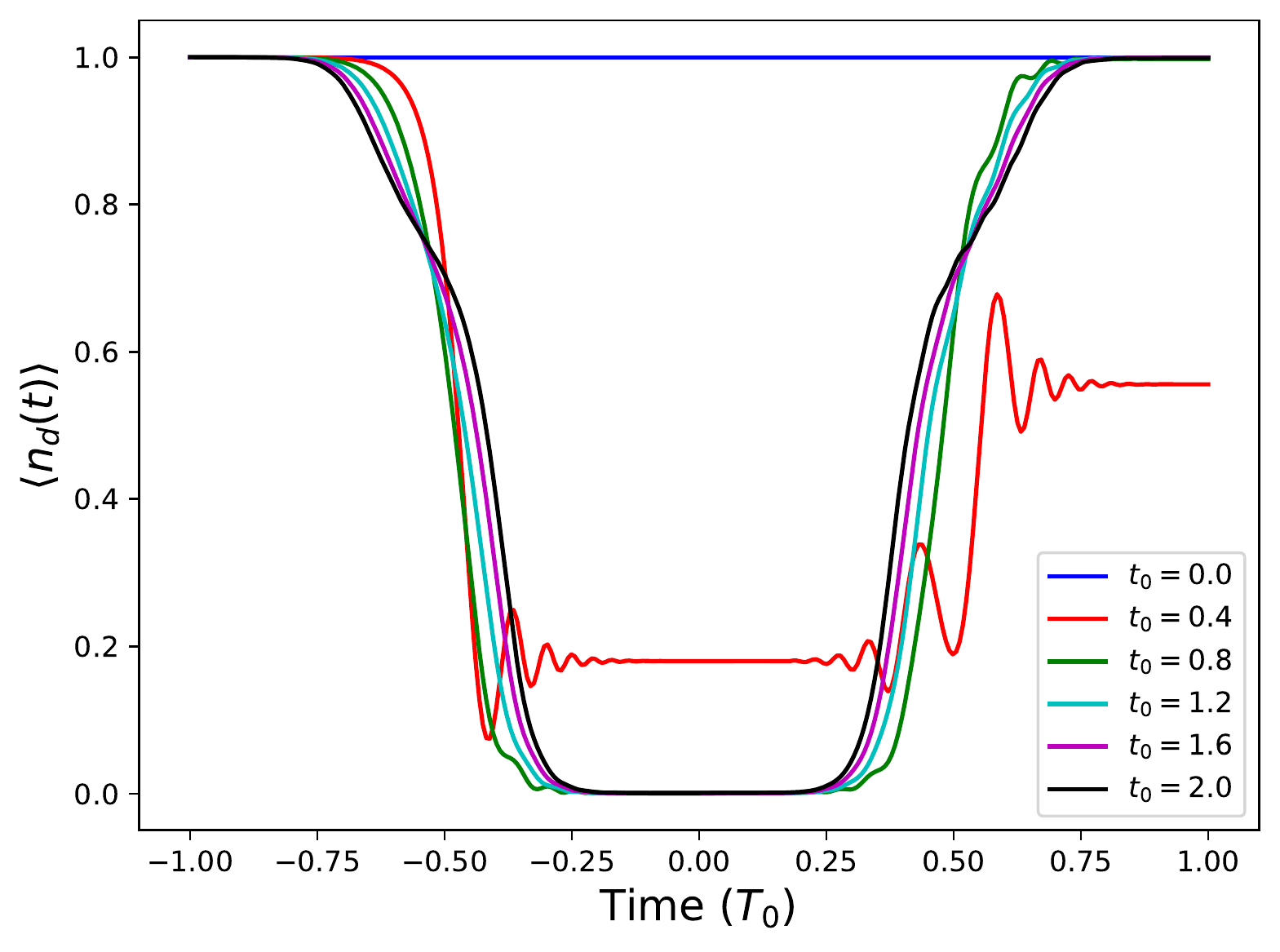}
\caption{Efficiency of LZ transition for the Parity Flip protocol at various coupling amplitude $t_0$ in units of 2$\epsilon_\text{max}$. Here, $(\sigma,\sigma')$ represents magnetic field orientation on the QD and initial wavefunction spin orientation.}
\label{fig:nanoLZ}
\end{figure}

\begin{figure}[htp]
\includegraphics[width=3.0in]{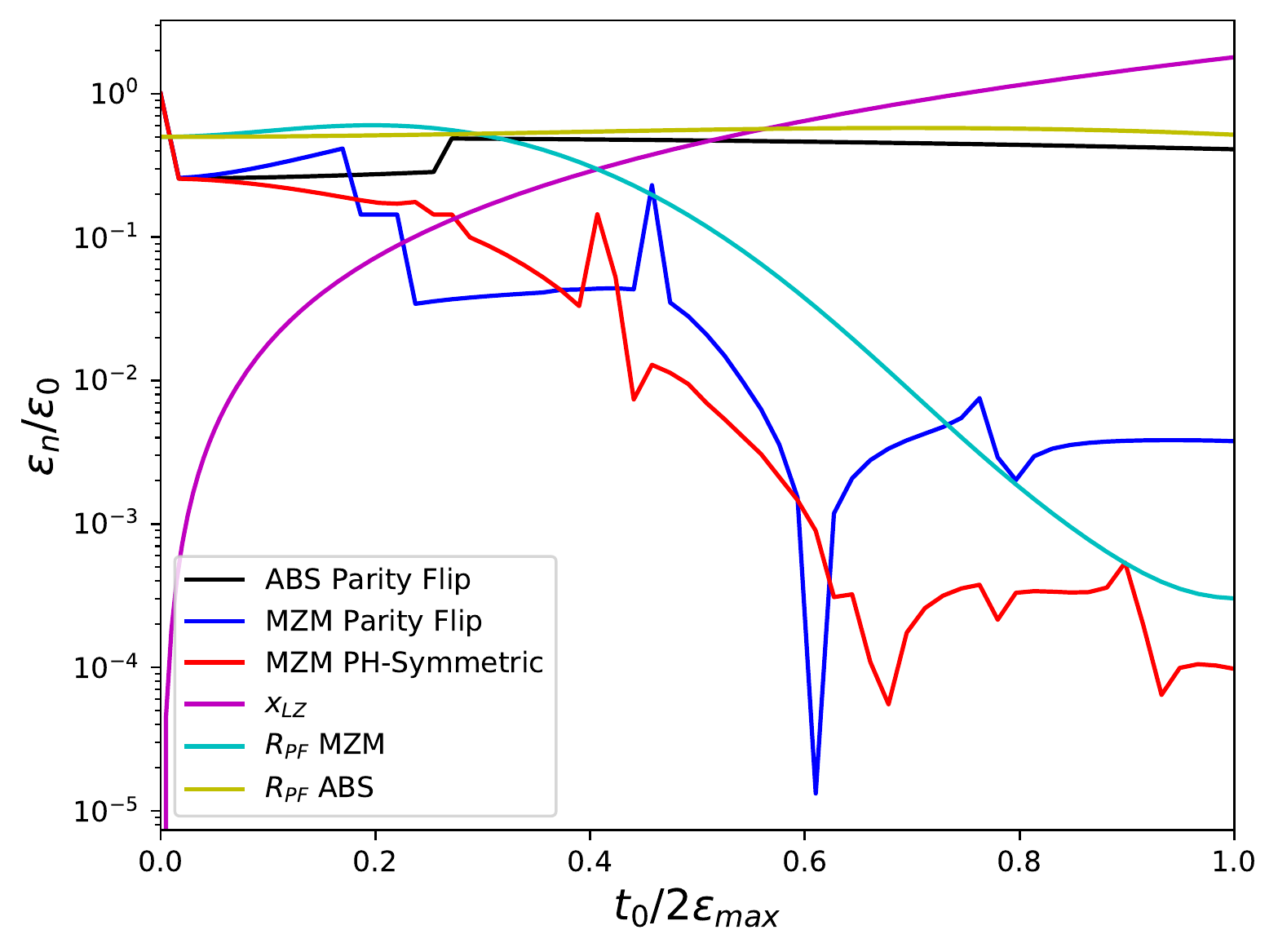}
\caption{Suppresed hybridization energy on realistic nanowire setup. $R_{\text{PF}}$ captures the efficacy of Landau Zener transitions in the parity-flip protocol. Smaller values are better. We see that the while the Landau Zener transition becomes more effective with increasing quantum dot-quantum wire tunnel coupling in the case of MZMs, it remains highly ineffective in the case of ABSs, clearly demonstrating that such tunneling is much more easily implemented with MZMs as opposed to ABSs.}
\label{fig:nanohyb}
\end{figure}

In Fig.~\ref{fig:nanopotentials}, we plot the potential on the quantum dot for the two spin states showing only one spin state within the bulk gap crossing the MZM energy. In Fig.~\ref{fig:nanoLZ}, we show the electron density on the quantum dot in time, showing the adiabaticity of successive Landau Zener transitions. In Fig.~\ref{fig:nanohyb}, we plot the renormalized hybridization energy for the various protocols implemented on the quantum wire with experimentally relevant parameters. 

\section{Conductance measurements}
\label{sec:conductance}

We now discuss the direct consequence of suppressed hybridization energy in tunneling conductance measurements. The detailed computation of the conductivity in the Floquet setting is discussed in App.~\ref{sec:computecond}. Here we present the essential details. 
To numerically compute normal to superconductor tunneling conductance in a time periodic system, we employ the Floquet Green's function method. The $k$-th harmonic component of the single-particle Green's function is defined as

\begin{equation}
    G^{(k)}(\epsilon)=\sum_{\alpha,k'}\frac{\ket{u_{\alpha,k'+k}}\bra{u^+_{\alpha,k'}}}{\epsilon-(\epsilon_\alpha+k'\Omega-i\gamma_\alpha)}
\end{equation}

where $\ket{u_{\alpha,k'}}$ and $\bra{u^+_{\alpha,k'}}$ are the $k'$-th harmonic components of the right and left Floquet eigenvectors~\cite{kohler2005driven} defined by
\begin{equation}
\begin{aligned}
\ket{u_{\alpha,k'}}=&\frac{1}{T}\int^T_0dt e^{ik'\Omega t} e^{i(\epsilon_\alpha-i\gamma_\alpha)t} \ket{u_\alpha(t)},  \\
\bra{u^+_{\alpha,k'}}=&\frac{1}{T}\int^T_0 e^{-ik'\Omega t} e^{-i(\epsilon_\alpha-i\gamma_\alpha)t} \bra{u^+_\alpha(t)}, 
\end{aligned}
\end{equation}
and $\alpha$ is an index enumerating distinct Floquet eigenvectors, and $\epsilon_\alpha - i \gamma_\alpha$ are the associated quasi-energies. Here $T$ is the full Floquet time period, and $\Omega=2\pi/T$ is the corresponding frequency. 

These eigenstates are obtained by numerically diagonalizing the Floquet time-evolution operator $U(t,0)$ given by
\be
    U(t,0)=\mathcal{T}_t\{ e^{ -i\int^t_0dt' h(t')-i d/2 }\}
\ee

where $h$ corresponds to QD and wire Hamiltonian matrix while $d$ corresponds to the lead self-energy term which enters as a non-Hermitian piece $i \frac{\Gamma_R}{2} c^\dagger_i c_i$ on wire sites that couple to the lead. The time-evolution is thus non-unitary and the left and right eigenvectors are distinct but together describe a bi-orthogonal basis satisfying $\sum_\alpha \ket{u_{\alpha} (t)} \bra{u^+_{\alpha} (t)} = \mathbb{1}$ at all times $t$. 

We note that the lead can only be coupled to the end of the wire opposite to the quantum dot. This is because this coupling generally inhibits successful implementation of the above proposed protocols as it can interfere with the process of electron tunneling between the dot and the wire. As a direct consequence, we can only use tunneling measurements in concert with the parity-flip and the PH-symmetric protocols---no wire end is left free to perform tunneling measurements in the left-right symmetric and composite protocols.

Finally, the time-averaged (over a Floquet period) conductance is given by the term 
\be
\sigma(\epsilon) =\frac{2e^2}{h}\sum_k \abs{F^{(k)}_{NN}(\epsilon)}^2\Gamma^2_R\\
\label{eq: conductance kitaev}
\ee
where $\abs{F^{(k)}_{NN}(\epsilon)}^2$ is the propagator corresponding to the conversion of an electron at site $N$ into a hole at site $N$ while absorbing $k$ units of photon energy $\Omega$. This result can be understood intuitively. Here we do not have any leakage from the wire except back into the leads. Thus, the only process that can contribute to the conductance is the Andreev reflection of electrons entering from the lead into the wire, and such reflections contribute a conductance of $2e^2/h$ multiplied by the amplitude for converting electrons (holes) into holes (electrons) at the wire site coupled to the leads. We note that we have assumed that the wire density of states is independent of the energy. For more detailed discussion of computing conductance in Floquet setting see Apps.~\ref{sec:computecond} for an analytical derivation of the result using Floquet theory, \ref{sec:condMZM} for the computation of this condutance for ideal MZMs and verifying that it indeed yields a $2e^2/h$ zero bias peak (ZBP) and \ref{sec: numdetcond} for numerical approach we take to compute the Green's functions. 

For the realistic nanowire, we have similar expression for the conductance, 
\begin{equation}
\label{eq: conductance realistic nanowire}
\begin{aligned}
    \sigma(\epsilon) =&\frac{2e^2}{h}\sum_{k\sigma\sigma'} |F^{(k\sigma\sigma')}_{NN}(\epsilon)|^2\Gamma^2_R, 
    \end{aligned}
\end{equation}
where we now additionally consider electrons of all possible spin orientations reflecting into all possible hole spin states. 

\begin{figure}[htp]
\includegraphics[width=3.2in]{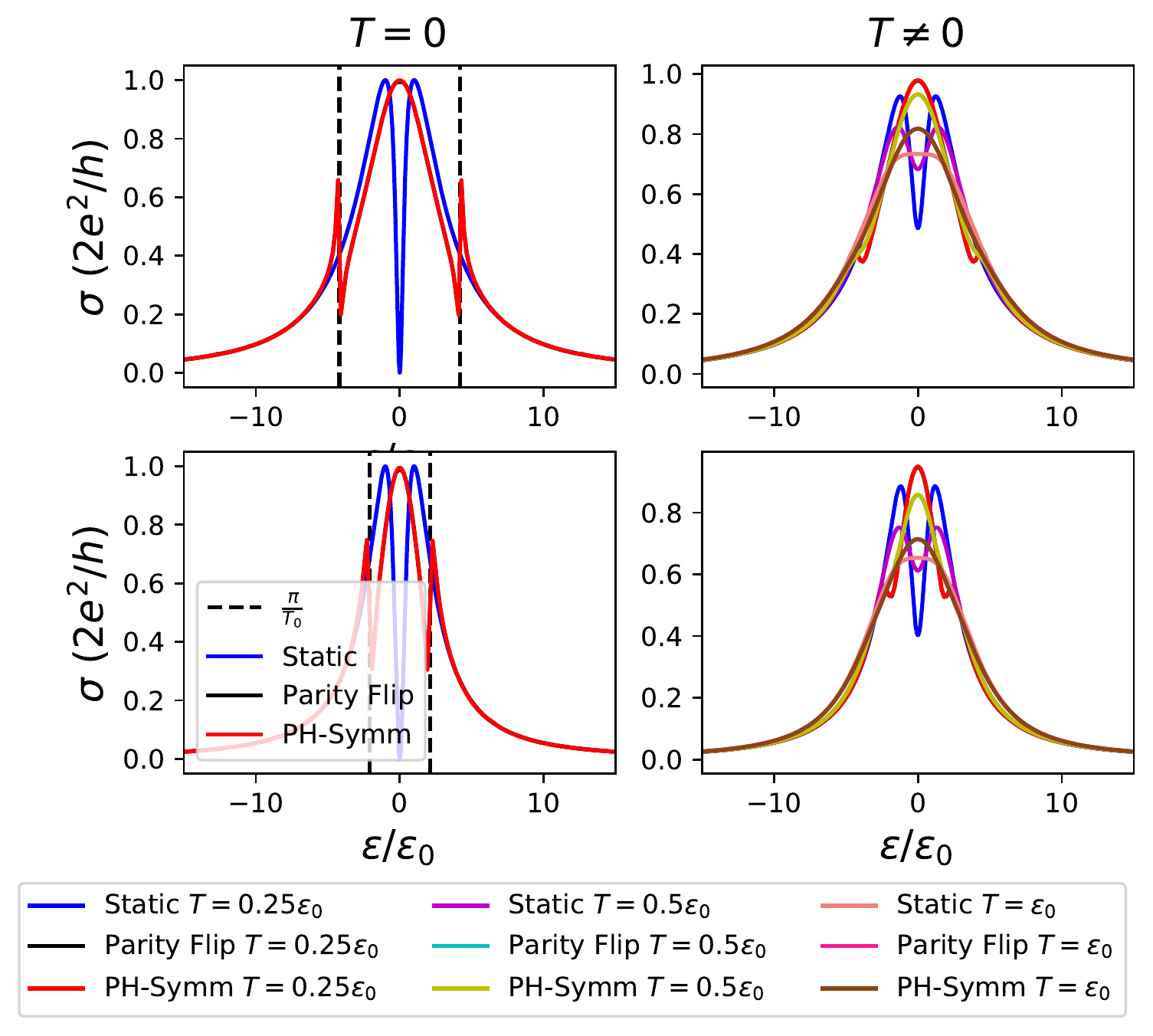}
\caption{Conductance peaks for static, Parity Flip, and PH-Symmetric protocols. The bottom legend is for plots on the right column. Note that conductance plots for Parity Flip and PH-Symmetric protocols almost lie on top of each other when viewed at the scale of the original hybridization energy between MZMs. The left column represents zero temperature while right column represents thermal broadening of conductance peaks at non-zero temperature. Top(Bottom) row is for $\mu=1.72$ ($\mu=1.8$) both with a right lead self energy of $25\epsilon_0$. Dashed lines correspond to the `photon energy' $\Omega=2\pi/T$ where $T$ corresponds to the full Floquet period for the protocol in question. The side peaks appear exactly at $\Omega$ with $T = 2T_0$ for the Parity Flip protocol and at $2\Omega$ with $T = 4 T_0$ for the PH-Symmetric protocol thereby lying on top of each other. These extra peaks are a direct consequence of number of mode oscillations during the protocol.}
\label{fig:cond_kitaev}
\end{figure}

\begin{figure}[htp]
\includegraphics[width=3.2in]{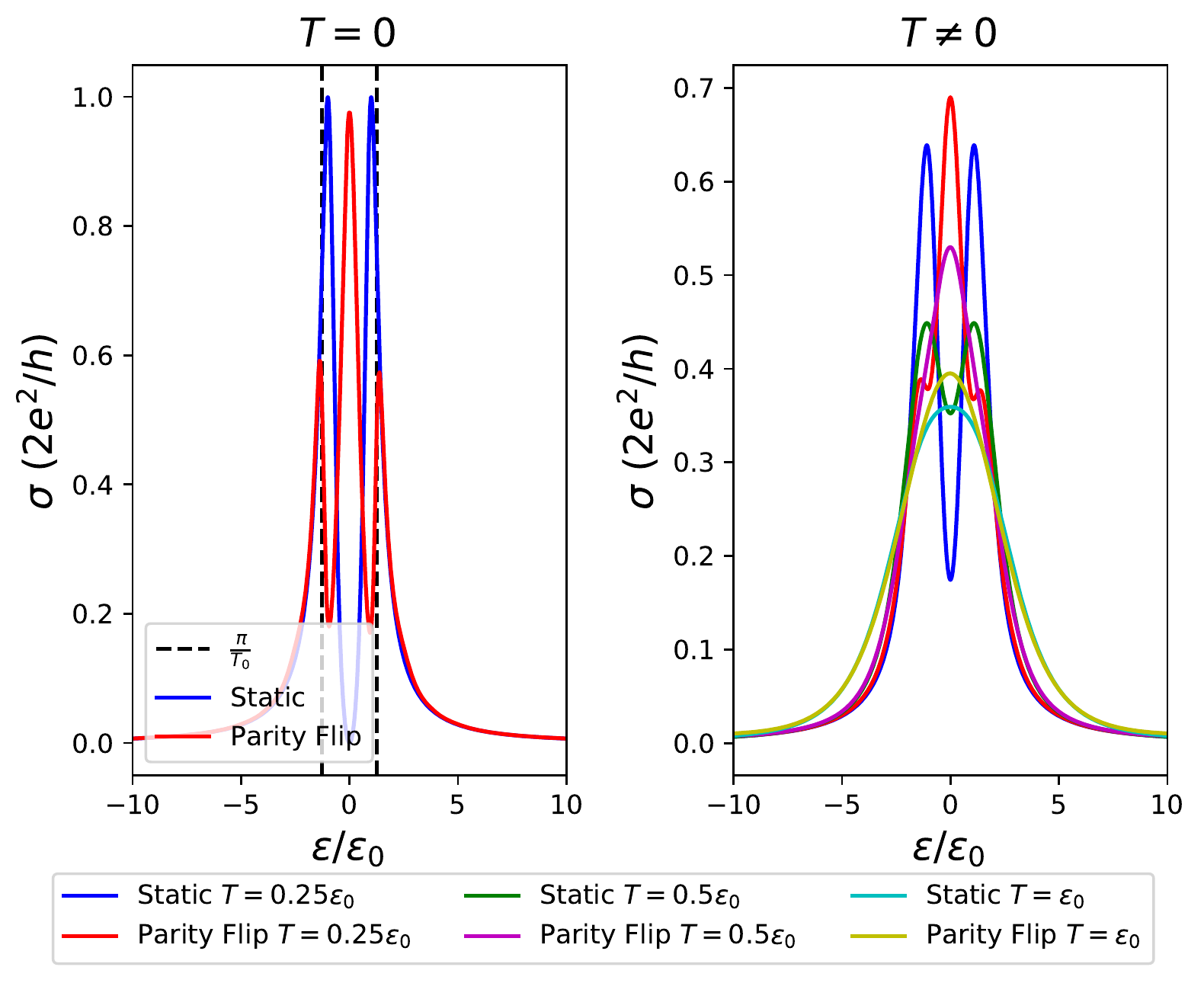}
\caption{Conductance peaks for static and Parity Flip protocols on a realistic semiconducting nanowire setup. The bottom legend is for plot on the right. The figure on the left represents zero temperature conductance while the figure on the right is for finite temperature. Here, we used the lead self energy to be $100\epsilon_0$. }
\label{fig:cond_nanowire}
\end{figure}

We also examine conductance at finite temperature which is given by
\begin{equation}
    \sigma(V,T\neq0)=\int d\epsilon \sigma(\epsilon)\frac{\beta}{4\cosh^2(\frac{\beta(\epsilon+V)}{2})} 
\end{equation}
where $\sigma(\epsilon)$ is the conductance at zero temperature defined in Eqs.~\eqref{eq: conductance kitaev} and \eqref{eq: conductance realistic nanowire}.

In general, we observe that the suppression of the hybridization energy due to the implementation of periodic driving is reflected in the shifting of conductance peak toward the zero bias; see Figs.~\ref{fig:cond_kitaev},\ref{fig:cond_nanowire}. In additional, one can also see a smaller peak at the frequency $\pi/T_0$, which is equivalent to the single `photon' energy in the case of the parity-flip protocol, and two-photon energy in the case of the PH-symmetric protocol. The origin of these peaks can be understood as follows. Recall that the Floquet eigenstates are composed primarily of the MZM operators $\gamma_L, \gamma_R$. In the Floquet setting, it is worth noting that these two operators sit at different frequencies identified by Floquet frequency folding---$\gamma_L$ sits at $\Omega$ or $2 \Omega$ as it flips sign once or twice, respectively, depending on the protocol in question. On the other hand, $\gamma_R$ does not oscillate in time and accrues phase at a rate given by the renormalized hybridization energy. Thus, when measuring the tunneling conductance from a lead coupled to the right end of the wire, electrons can tunnel into the wire jumping into, approximately, $\gamma_R$, at the reduced hybridization energy, or $\gamma_L$ at the reduced hybridization energy plus the one (or two) photon energy. These peaks thus correspond to a smaller amplitude but nevertheless show up as the overlap between the MZMs in these systems is not irrelevant. 

Note that while this shift of conductance peak towards zero bias is visible at zero temperature, it is quickly broadened at finite temperature and it becomes hard to distinguish, from tunneling measurements, the effect of applying the protocols we propose. 

Finally, we note that we have not studied these measurements in the presence of ABSs---this is because it is impossible to simultaneously place a quantum dot to drive the system and couple the lead to the same end to measure the conductance (and placing the tunneling lead on the other end of the wire where ABSs are not present is not sensible either). However, it is worth noting that if ABSs exist on both ends of the wire, then conductance measurements on the other end of the wire will show no change in the presence of ABSs, as opposed to when MZMs are present. This is in spirit akin to two-terminal conductance measurements where the true non-local nature of the MZMs becomes apparent and distinguishes them from ABSs.  


\section{Qubit coherence measurements}
\label{sec:qubitcoherence}

In this section, we discuss the consequences of the Floquet protocols for the coherence of the quantum bits constructed out of MZM or ABS qubits. Our approach is inspired by Ref.~\cite{mishmashdephasing} which studied the dynamics of the $\avg{\sigma_x (t) \sigma_x (0) }$ autocorrelator of a putative tetron qubit in the presence of MZMs and ABSs. Here $\sigma_x$ is simply the Pauli-X operator implemented on the MZM qubit subspace---it shows oscillations at a characteristic frequency given by the hybridization energy of the MZMs on the wire, and its decay, obtained by averaging over multiple experimental runs in the presence of dephasing noise yields the $T_2$ coherence time of the qubit (assuming the quasi-particle poisoning rate is much smaller). For numerical convenience, we will simulate such correlators for an MZM qubit on a single quantum wire. We discuss the adaptation of our ideas to the tetron qubit in the next section. 

The Floquet protocols are implemented using the same parameters as in Secs.~\ref{sec:hybenergy} and~\ref{sec:conductance}. We study the relaxation of this autocorrelator in the presence of dephasing noise with a $1/f$ noise spectrum. In particular, we consider global fluctuations of the chemical potential $\mu (t)$ with mean $\avg{\mu(t)} = 0 $, and variance $\avg{(\mu(t)-\bar{\mu})^2} = \delta \mu^2$, and a frequency spectrum of the form 
\begin{align}
S_{\mu} (f) &=  \frac{A(f_\text{max},f_\text{min})}{f}, \; \; \;  f_{\text{min}} < f < f_{\text{max}}. 
\label{eq:noisespectrum}
\end{align}

which is non-zero in the frequency regime specified. Note that the spectrum is produced by adding Fourier modes with the appropriate weight [given by Eq.~(\ref{eq:noisespectrum})] and then setting the amplitude $A$ by forcing the time-dependent signal to have a given mean and variance; this is particularly important for $1/f$ noise, where the low-frequency part in particular is effectively static over the course of the simulation and may not have a mean that is precisely zero unless it is explicitly corrected for. 

This spectral form is natural for electromagnetic noise~\cite{paladino20141fnoise} (whose origin can be from multiple sources, including two-level systems trapped at the interface between the wire and the substrate, among others~\cite{agarwal1fnoiseheisenberg,burnett2014evidence,agarwalpolaron}) and is suggested to be a significant impediment to realizing longer coherence times in MZM qubits~\cite{knapp2018dephasing}. We note that the lower frequency cutoff can be remarkably low in many physical systems~\cite{antonclarke1fnoie}---it is thus effectively determined by the duration of the experiment. The UV cut-off depends on microscopic details and we therefore study the effect of the variation of this cut-off on the dephasing of the qubits. 

\subsection{Impact of dephasing noise on the quantum dot}

In our simulations, we also allow for some local potential noise on the QD with the same spectral form as the chemical potential variations on the wire but with potentially a different magnitude. Naively, it may appear that our new Floquet MZM qubits are limited by the coherence times imposed by fluctuations on the QD as the electron sits on the QD at intermediate times; this would be disastrous because a major benefit of working with MZM qubits in the first place is to make the qubits immune to dephasing at the hardware level.  

Fortunately, a close inspection of Fig.~\ref{fig:protocols} reveals that in fact, the state of the quantum dot is identical and independent of the parity of the wire at all times. This implies that dephasing noise on the quantum dot should not deteriorate qubit coherence provided it is not too strong. We further examine this aspect below. 

Note that the ratio of the dephasing rate of a putative MZM qubit, $\Gamma_{\text{MZM}}$, and a quantum dot qubit, $\Gamma_{\text{QD}}$, is given by the ratio of the relative sensitivity of qubit energies to the change in the chemical potential. Thus, 

\begin{align}
   \frac{\Gamma_{\text{MZM}}}{\Gamma_{\text{QD}}} = \left( \frac{\d{\epsilon_{0}}{\mu}}{\d{ \epsilon_{D}}{\mu}} \right)^2 \sim \beta^2
\end{align}

where $\beta \equiv e^{-L/\xi}$ is the hallmark suppression factor that emerges from the non-locality of the MZM qubit. We now show that the renormalized hybridization energy of the Floquet-MZM qubit, estimated in Eq.~\ref{eq:Epm} for the parity-flip protocol, also depends only through a factor of $\beta \text{d} \epsilon_D / \text{d} \mu$ on the fluctuations of the energy of the quantum dot due to noise.

In particular, we again note that the phase \emph{difference} accrued, $\propto E_+ - E_-$ between the eigenstates in the two distinct parity sectors does not depend on the energy of the quantum dot except at times when the tunneling matrix element between the quantum dot and the quantum wire is relevant, and $\epsilon_D (t) \sim \epsilon_0$. Mathematically, away from the tunneling regime, $\epsilon_D$ dominates the instantaneous energy for both eigenstates and cancels in the energy difference. Physically, this is because the state of the quantum dot is identical in these regimes in the Floquet eigenstate. 

Next, it is straightforward to ascertain that

\begin{align}
    \d{(E_+-E_-)}{\mu} &= \d{\epsilon_D}{\mu} \Bigg[ \frac{- \epsilon_0 + \epsilon_D /2 }{\sqrt{(\epsilon_0 - \epsilon_D/2)^2 + \frac{t^2_0}{\xi} \abs{1 + i \beta}^2 }} \nonumber \\
    &- \frac{\epsilon_0 + \epsilon_D/2}{\sqrt{(-\epsilon_0 - \epsilon_D/2)^2 + \frac{t^2_0}{\xi} \abs{1 + i \beta}^2 }} \Bigg] \sim \beta \d{\epsilon_D}{\mu} 
    \label{eq:dotnoise}
\end{align}

where it is easy to check that the bracketed term vanishes in the limit $\beta \rightarrow 0$ (note that $\epsilon_0 \propto \beta$). Thus, we note that the effective noise on the quantum dot that enters the renormalization hybridization energy is weakened signficantly, and by the same factor $\beta$ that makes the MZM qubit stable to noise. We thus anticipate that noise on the quantum dot does not significantly impact the protocols we propose. Numerical examination of the Floquet MZM qubit's coherence time shows that it decreases by a factor of $2$ in the presence of noise on the quantum dot as opposed to without any noise on the quantum dot. This is in complete contrast to the naive expectation that noise of the same magnitude on the quantum dot as on the wire should instantly decohere the Floquet MZM qubit, but instead follows logically from the result of Eq.~(\ref{eq:dotnoise}) which shows that the noise on the quantum dot effects the coherence of the Floquet MZM qubit suppressed by a small factor of $\beta^2$. 

Note finally that the above discussion assumes that the efficacy of the Landau-Zener transitions is not significantly impacted by the presence of noise on the quantum dot. This is justified when the noise frequency is lower than the inverse of timescale on which the quantum dot's energy is comparable to the hybridization energy of the MZMs and the variation of the chemical potential due to noise, $\delta \mu$, is small compared to $\epsilon_{\text{max}}$, the scale on which the quantum dot's energy is varied. In this case, a slowly changing shift of the potential of the quantum dot can change the precise time at which the Landau-Zener transition occurs, but it cannot destroy its efficacy when it does occur. 

\subsection{Numerical Results}

\begin{center}
\begin{figure}[htp]
\includegraphics[width=3.3in]{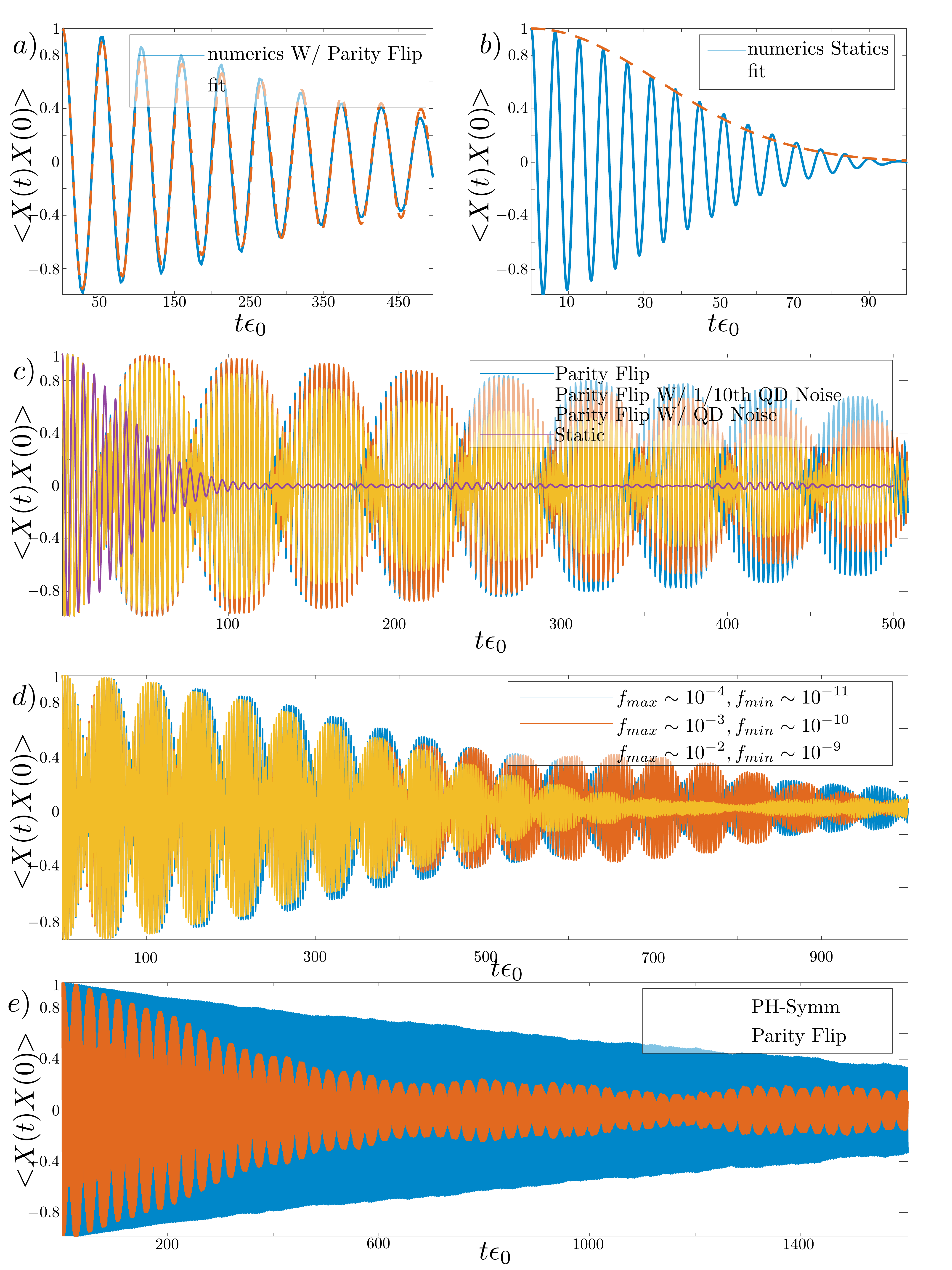}
\caption{Autocorrelator averaged over 30 samples of 1/f noise for the Parity Flip protocol, PH-symmetric protocol and the static case. Here we use $t0/2\epsilon_{\text{max}} = 0.5$, with the remaining parameters as noted in the previous section. The noise on the wire has been chosen with $f_{max} = 10^{-3}$ and $\delta \mu = 0.002$ for all plots except for (c) where $f_{max}$ is varied. (a),(b) The autocorrelator is fit to $e^{- t/T^{p}_{2}}\cos(\epsilon_{n} t)$ for the parity-flip protocol and $e^{- (t/T_2)^2}$ for the static case where we anticipate Gaussian decay of correlations due to an unsupressed $1/f$ noise spectrum. We find $T^{p}_2 = 2.6 \times 10^{5}$,  $T_2 = 2.4083 \times 10^{4}$ and $\epsilon_{n} = 2.3 \times 10^{-4}$. The renormalized hybridization energy and the coherence times both show an order of magntiude improvement over their static counterparts. (c) The autocorrelator is computed with varying noise magnitude on the QD equal to that on the wire, an order of magnitude smaller, and in the absence of noise on the QD. Increasing noise on the QD reduces the coherence of the MZM qubit but the effect is not as dramatic as naively expected (see main text for  discussion). (d) The maximum and minimum cut-off frequencies of the 1/f noise spectrum are varied. (e) Comparison between the Parity flip and PH-symmetric protocol with the latter showing a further order of magnitude improvement in coherence.}
\label{fig:kitaevcoherence1}
\end{figure}
\end{center}

Figs.~\ref{fig:kitaevcoherence1} show the decoherence of the MZM qubit in the presence of dephasing noise with and without the application of the protocol. It is apparent that the parity-flip protocol can increase the $T_2$ coherence time of the qubit by at least an order of magnitude for experimentally relevant parameters. The autocorrelator can be used to extract the renormalized hybridization energy, as we show in Fig.~\ref{fig:kitaevcoherence1} (a), which shows the autocorrelator sampled at the drive period $2T_0$ of the parity-flip protocol. Turning off the noise on the quantum dot leads to an improvement in the Floquet MZM qubit's coherence time but only by a modest factor of $2$, in accordance with the intuition developed above that the amplitude of noise on the quantum dot does not directly impact the coherence of the Floquet qubit; see Fig.~\ref{fig:kitaevcoherence1} (c). The Floquet qubit's coherence times appear to become worse as the high frequency cut-off exceeds the inverse driving timescale as such high frequency noise is not suppressed by the protocol we implement; nevertheless, for a maximum frequency $f_{\text{max}}\sim 10^{-2}$ which is an order of magnitude larger than the MZM qubit's energy $10 \epsilon_0$, shows no significant worsening, presumably because the high frequency cut off has limited weight in the noise spectrum---see Fig.~\ref{fig:kitaevcoherence1} (d). The PH-symmetric protocol appears to result in further improvement over the parity-flip protocol, as seen in Fig.~\ref{fig:kitaevcoherence1} (e). It is worth noting that some of the plots exhibit beating-like behavior---we note the period is half the period set by oscillations of the qubit's autocorrelator at the renormalized hybridization energy, and is absent when the autocorrelator is plotted for times that are multiples of the drive period, $2T_0$. This is evident from comparing the results of Fig.~\ref{fig:kitaevcoherence1} (a) where we show only results at multiples of the drive period, and the plots below where we sample within each drive period. 

\begin{center}
\begin{figure}[htp]
\includegraphics[width=3.3in]{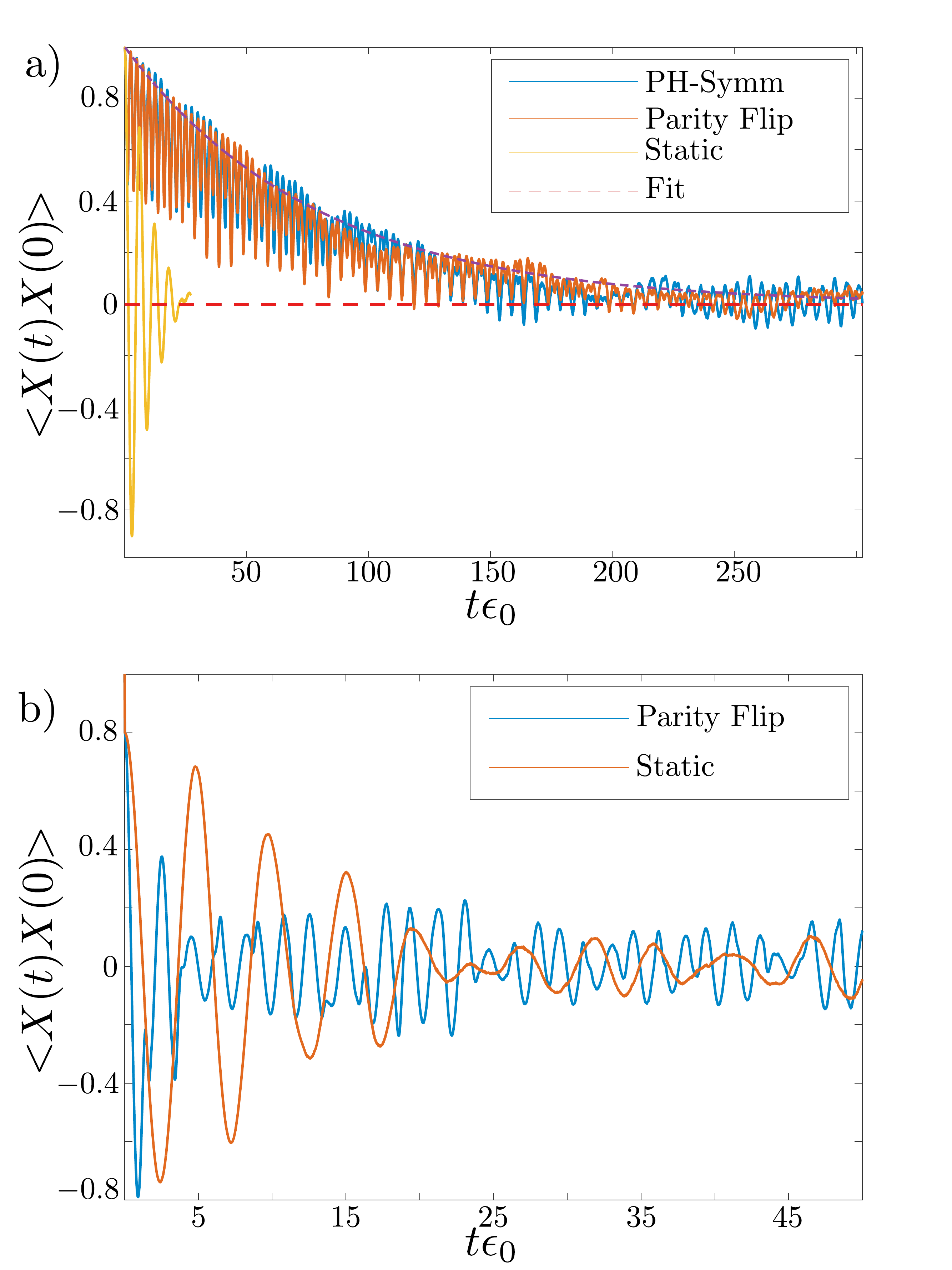}
\caption{Autocorrelator averaged over 30 samples of 1/f noise for the Parity Flip protocol and the static case in the case of the realistic nanowire in the presence of MZM (a) and ABS (b). The coherence is generally improved for MZM based qubit, but reduced for ABS based qubit. Here we use $t0/2\epsilon_{\text{max}} = 1$, with the remaining parameters as noted in the previous section. The noise on the wire has been chosen with $f_{max} = 10^{-3}  \text{meV}/\hbar \approx 1 \text{KHz}$ and $\delta \mu = 0.002$ meV. The noise on the QD is the same as that on the wire with its magnitude reduced by a factor of ten. (a) The autocorrelator is fit to $e^{- t/T^{p}_{2}}$ for the parity-flip case. We find $T^{p}_2 = 2.9 \times 10^{4} \hbar/\text{meV}$.}
\label{fig:MZMABScoherence1}
\end{figure}
\end{center}

Finally, we note that  simulations on the realistic nanowire yield similar results; see Fig.~\ref{fig:MZMABScoherence1} (a)---in these simulations, we assume the noise on the quantum dot to be $1/10^{\text{th}}$ of the noise magnitude on the quantum wire. Both PH-symmetric and parity-flip protocols equally improve the coherence of the qubit by an order of magnitude in this case. Unlike the Kitaev case, the PH-symmetric protocol does not seem to improve the coherence over the parity-flip protocol, but we anticipate this occurs due to how adiabatic the Landau Zener transitions are for the parameters used in simulations in the two cases as these strongly affect the performance of the PH-symmetric protocol. 

Finally, we study the effect of such protocols on a noisy wire harboring ABSs. As mentioned above, we anticipate that the tunneling coupling significantly affects the energy of the ABSs which makes it very unlikely to effect perfectly adiabatic Landau Zener transitions. This implies that Floquet eigenstates differ significantly in their spatial form compared to the original ABS states. As a result, the autocorrelator should oscillate with multiple frequencies (whose amplitude is determined by the overlap of the static ABSs on the Floquet eigenstates), an expectation which is verified in simulations. Moreover, in the presence of noise on the quantum dot, these oscillations die very rapidly, showing in fact worsening of coherence times as compared to the case where the protocol is not implemented. See Fig.~\ref{fig:MZMABScoherence1} (b) for details. 

The above conclusions suggest that measurements of such autocorrelators can provide a very clear dynamical signature that distinguishes ABSs from MZMs. Of course, it would be challenging to implement such an experiment as it requires observing the coherence of an MZM qubit in its two different parity states. In the next section, based on our numerical simulations thus far, we propose a concrete experiment based on a tetron qubit~\cite{plugge2017majorana,karzig2017scalable} where a variation of our protocol may be implemented and be used to clearly distinguish ABSs from MZMs.  

\begin{center}
\begin{figure}[htp]
\includegraphics[width=3.2in]{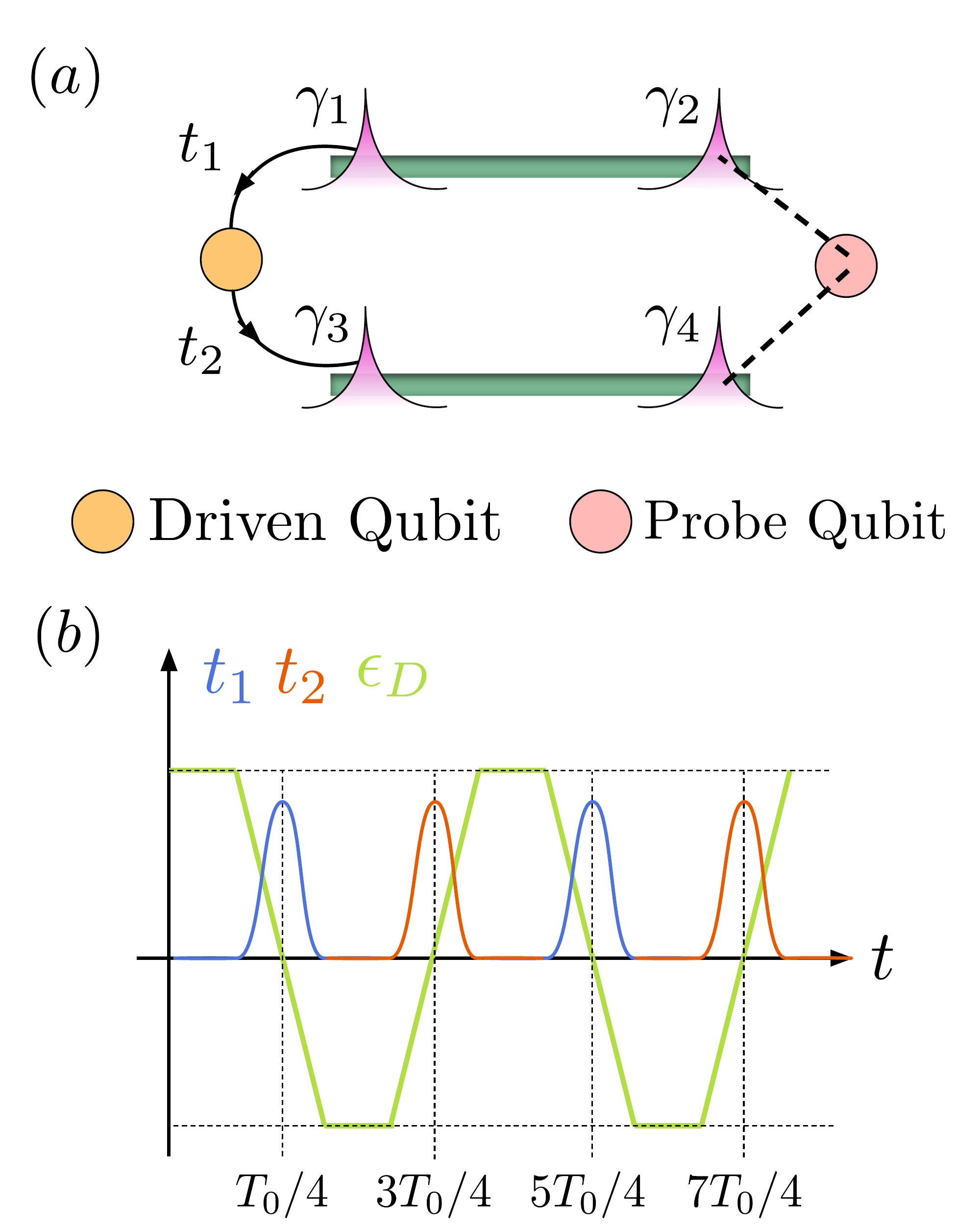}
\caption{(a) The experimental setup---the left end of the quantum wires are coupled to a quantum dot with time-dependent couplings $t_1$ and $t_2$ respectively and on which a potential $\epsilon_D (t)$ is applied. A probe quantum dot on the right end of the wires can be used to initialize the qubit and measure the $\sigma_x$ operator. (b) The time-dependent potential $\epsilon_D (t)$ and couplings $t_1 (t), t_2 (t)$.}
\label{fig:tetron}
\end{figure}
\end{center}

\section{Experimental Implementation in Tetron Qubits to distinguish between Andreev bound states and Majoranas}
\label{sec:exptetron}

Although thus far we have outlined how various Floquet protocols may be implemented on a single nanowire with the aid of a QD, it is clear that the protocols can be easily generalized to the case of the tetron qubit. The tetron is composed of two nanowires each supporting MZMs; see Fig.~\ref{fig:tetron}. The combined parity of the two wire system is fixed, so that $P = - \gamma_1 \gamma_2 \gamma_3 \gamma_4 = 1$ (say) is fixed. The Pauli operators for the tetron qubit can be then identified as $\sigma_z = - i \gamma_1 \gamma_2$, $\sigma_x = -i \gamma_1 \gamma_3 = i \gamma_2 \gamma_4 $. If the hybridization energy of the MZMs on the two wires are given by $\epsilon_{12}, \epsilon_{34}$, then the energy splitting of the qubit is given by the difference $\epsilon_{12} (t) - \epsilon_{34} (t) $, where both the energies acquire some time dependence due to electromagnetic noise. Then, a simple extension of the PH-symmetric protocol to the tetron qubit case is to use a QD on the left end of these wires, see Fig.~\ref{fig:tetron} where we denote it as `driven qubit', to effect transitions from the top and bottom wires at times $T_0/4, T_0 + T_0/4$ and $3T_0/4, T_0 + 3T_0/4$ respectively. This ensures both wires spend equal time in both parity sectors \emph{and} both wires engage in conventional and Andreev tunneling which ensures that the renormalized effective qubit energy splitting is reduced in the same way as in the PH-symmetric protocol implemented for a single nanowire. We thus anticipate that the reduction will only be limited by non-adiabaticity during Landau-Zener transitions.   

The measurement of the autocorrelator $\avg{\sigma_x (t) \sigma_x (0)}$ can then be performed by i) initializing the system in a state with definite $\avg{\sigma_x (0)} = \avg{i \gamma_2 \gamma_4}$, say by adiabatically switching on a large tunnel coupling using a probe quantum dot on the other end of the wires, and ii) measuring $\avg{\sigma_x (t)}$ using this probe quantum dot at a later time $t$. Based on our numerical and analytical observations above, we anticipate  that this measurement should reveal a clear distinction between MZMs and ABSs. In the former case, one should see a decrease in the frequency of oscillations as compared to measurements done in the absence of the Floquet protocol, accompanied by an increase in the coherence time of the qubit by an order of magnitude or more. In the latter case of ABSs, one should see oscillations at no definite frequency, and which relax on a timescale similar or worse to the coherence time of the qubit measured without the application of the protocol. 

\section{Conclusions and Discussions}
\label{sec:conclusions}

We now summarize our main findings and provide context to our results. Although much theoretical and experimental progress has been made towards the realization of Majorana zero modes (MZMs), there remain important challenges that hinder further progress. First, various engineering challenges have meant that the MZMs realized in experiments are not truly isolated as one would desire for the purposes of realizing the full potential of MZMs towards quantum computing. Second, these imperfect topological qubits are expected to be sufficiently impacted by electromagnetic $1/f$ dephasing noise which potentially limits their application as qubits. Finally, it has also been suggested that trivial Andreev bound states (ABSs) can remain pinned robustly near zero energy making it hard to distinguish them for true topological excitations like MZMs in local tunneling measurements. 

In this work, we have discussed several Floquet protocols, for both a single quantum nanowire harboring MZMs, and tetron qubits composed of two such wires, to address the challenges posed above. Crucially, the basis for these protocols is a property that is unique to MZMs, even if they are imperfect---double exchanging MZMs results in a robust negative sign which can then be used to implement dynamical decoupling techniques to reduce the hybridization between MZMs and the deleterious effect of both dephasing noise and quasiparticle poisoning. When the Majorana operators constituting the bound state are not physically separated, even if they only hybridize very weakly as in the case of an ABS at very low energy, the above exchanges cannot be successfully implemented. As a result, while the coherence of a topological qubit improves dramatically when such protocols are implemented, the coherence of a trivial ABS qubit in fact degrades (marked by a lack of oscillations in time-dependent correlations of the qubit, and a rapid decay). Importantly, this distinguishing signature can be observed without changing external parameters, or requiring multiple nanowires of different lengths. (It is worth noting that severals works have proposed ways to distinguish ABSs and MZMs using sharp edge potentials in the expectation that the former respond much more strongly to such a perturbation; see Refs.~\cite{deng2018nonlocality,clarkeABSMZM,Pradaquantumdot2017,liu2018distinguishing}. Our work utilizes this principle as well but additionally proposes to use this to improve coherences of MZM qubits.)    

While double exchanging MZMs may appear to be a daunting task, Ref.~\cite{agarwalmajoranadraiding} showed that this can be achieved, with the same robustness as physically braiding MZMs, using a quantum dot coupled to the nanowire, and driving it to force adiabatic transfer of charge between the quantum dot and the wire. Physically, this leads to the rapid and periodic flipping of the parity on the nanowire, which is mathematically equivalent to changing the sign of the hybridization term in the MZM Hamiltonian. This leads to the suppression of both the static hybridization between the MZMs and low-frequency dephasing noise. What is peculiar about MZMs, however, is that sign flip is effected by a change of the sign of the Majorana mode adjacent to the quantum dot, which should suppress quasiparticle tunneling on that end of the nanowire. 

The parity-flip protocol proposed in Ref.~\cite{agarwalmajoranadraiding} can be bettered by more careful accounting of phases accrued in the time-evolution of the two parity states of the nanowire. We introduced the PH-symmetric protocol whose ability to suppress the hybridization between end MZMs appears to be limited only by the adiabaticity of the Landau-Zener transitions. Additionally a left-right symmetric protocol was introduced employing a quantum dot on each edge of the nanowire, in order to suppress quasiparticle poisoning on both ends of the wire. The positive features of both these protocols can be combined into a `composite' protocol. Finally, we also introduced an analog of the PH-symmetric protocol for a tetron qubit consisting of two such nanowires, see Fig.~\ref{fig:tetron}. Simulations of these protocols were used to provide credence to the ideas described above. 

\section{Acknowledgements}

The authors acknowledge useful discussions with Jukka Varkynyn and Yuval Oreg. KA acknowledges previous collaboration on related work with Ivar Martin. KA and TPB acknowledge the hospitality of the Aspen Center for Physics where parts of the ideas presented here where conceived. KA ackowledges funding from NSERC, FRQNT and a Tomlinson Scholar award. TPB acknowledges support from NSERC and FRQNT.   

\appendix

\section{Computing conductance in the Floquet setting}
\label{sec:computecond}

Here we detail the computation of the tunneling conductance (as well as non-local conductance) in the Floquet setting. We follow Ref.~\cite{kohler2005driven} that studies transport in periodically driven quantum nanowires and generalize their discussion to superconducting systems.  

We will discuss the computation for a spinless system; the generalization to the spinful case is immediate and results will provided after. The generic Hamiltonian characterizing a spinless nanowire is given by

\begin{align}
H_{\text{wire}} (t) &= \sum_{n,n'} H_{n,n'} (t) c^\dagger_n c_{n'} + \nonumber \\
&\frac{1}{2} \sum_{n,n'} \left[ \Delta_{n,n'} (t) c^\dagger_n c^\dagger_{n'} + \Delta^*_{n,n'} (t) c_{n'} c_n\right]
\end{align}

where, by Hermiticity, $\Delta_{n,n'} = - \Delta_{n,n'}$ and $H_{n,n'} = H^*_{n',n}$, and $n$ characterize lattice sites. 

Left and right leads connecting to this wire are characterized by modes $c_{L q}$ and $c_{R q}$ respectively. The Hamiltonian of the leads and the contacts is given by

\begin{align}
H_{\text{leads}} &= \sum_q \epsilon_q \left( c^\dagger_{L q} c_{L q} + c^\dagger_{R q} c_{R q} \right) \nonumber \\
H_{\text{contacts}} &= \sum_q \left( V_{L q} c^\dagger_{L q} c_1 + V_{R q} c^\dagger_{R q} c_N \right) + \text{h.c.}
\end{align}

where $n = N$ characterizes the last site of the nanowire. 

The equation of motion of the lead fermions is given by $\dot{c}_{l q} = - i \epsilon_q c_{l q } - i V_{l q} c_{n_l}$ where $l = L, R$ and $n_L = 1, n_R = N$. This can be formally solved to yield

\be
c_{l q} (t) = c_{l q} (t_0) e^{-i \epsilon_q (t - t_0)} - i \int_0^{t-t_0} V_{l q} c_{n_l} (t - \tau) e^{-i \epsilon_q \tau} d \tau
\label{eq:app1}
\ee

For wire modes, the equation of motion can be similarly computed, and inserting Eq.~(\ref{eq:app1}), we find

\begin{align}
\dot{c}_{n_l} &= -i \sum_n H_{n_l n} c_n - i \sum_n \Delta_{n_l n} c^\dagger_n \nonumber \\
&- \int_0^\infty \Gamma_l (\tau) c_{n_l} (t-\tau) d \tau + \xi (t) \nonumber \\
\dot{c}_{n \neq n_l} &= - i \sum_{n'} H_{n n'} c_{n'} - i \sum_{n'} \Delta_{n n'} c^\dagger_{n'}. 
\label{eq:app2}
\end{align}

where we took $t_0 \rightarrow - \infty$, and defined 
\begin{align}
\Gamma_l (\tau) &= \int_0^{\infty} \sum_q \abs{V_{l q}}^2 e^{-i \epsilon_q \tau} d \tau \nonumber \\
\xi_l (t) &= - i \sum_q V^*_{l q} c_{l q} (t_0) e^{- i \epsilon_q (t - t_0)}. 
\end{align}

which serve as the dissipation and fluctuation in a Langevin equation for the wire modes coupled to the leads. The fluctuations can be assumed to have a Gaussian thermal distribution (assuming the leads are always in equilibrium, at a definite temperature). The correlations are given by $\avg{\xi^\dagger_l (\epsilon) \xi_{l'} (\epsilon')} = 2\pi \delta( \epsilon - \epsilon') \delta_{l l'} f_l (\epsilon) \Gamma_l (\epsilon)$, where $f_l (\epsilon)$ is the Fermi-Dirac function, denoting the occupation number of electrons in the lead $l$ at energy $\epsilon$. 

It is useful to organize these equations by defining Green's function matrices


\begin{equation}
\begin{aligned}
 G_{n n'} (t,t') &\equiv  - i \Theta ( t - t') \langle c_n (t), c^\dagger_{n'} (t') \rangle,\\
 F_{n n'} (t,t') &\equiv -i \Theta (t - t') \langle c^\dagger_n (t), c^\dagger_{n'} (t') \rangle\\
 g \equiv \begin{pmatrix} G & -F^* \\ F & - G^*  \end{pmatrix} &, \; h \equiv \begin{pmatrix} H & \Delta \\ -\Delta^* & - H^* \end{pmatrix}, \; d \equiv \begin{pmatrix} \Gamma & 0 \\ 0 & \Gamma^* \end{pmatrix}
\end{aligned}
\end{equation}

where $\Gamma$ has been generalized to imply the matrix $\Gamma_{nn'} = \delta_{n n'} \left( \delta_{n 1} \Gamma_L + \delta_{n N} \Gamma_R \right)$. 

Eqs.~(\ref{eq:app2}) can be solving in terms of the Green's function which satisfies the equation

\be
\d{g}{t} = -i \delta (t - t') \mathbb{1} - i h g - \int_0^\infty d\tau d(\tau) g ( t- \tau). 
\label{eq:gapp}
\ee

leading to 

\begin{align}
    c_n (t) &= i \hbar \sum_l \int_0^\infty d\tau \; g_{n n'} (t, t-\tau) \xi_{n'} (t-\tau)  \nonumber \\
    &= \frac{i\hbar}{2\pi} \sum_l \int_{-\infty}^{\infty} d \epsilon \; g_{n n'}(t, \epsilon) \xi_{n'} (\epsilon) e^{- i \epsilon t},  \nonumber \\
    \xi_n &\equiv [\xi_L , 0, ..., 0, \xi_R, \xi^*_L, 0, ..., 0, \xi^*_R]^T. 
    \label{eq:app3}
\end{align}

In the above we performed a Fourier transform of the time-difference between the two time arguments of the Green's function, using $g (t, \epsilon) = \int_{-\infty}^{\infty} d\tau e^{i \epsilon \tau} g(t, t-\tau)$. Note that, in our Floquet setting, the $g(t, \epsilon)$ can be further expanded in terms of the harmonic components in the Floquet period $T$---

\begin{align}
g(t, \epsilon) &= \sum_k g^{(k)} (\epsilon) e^{-ik \Omega t} \nonumber \\
g^{(k)} (\epsilon) &= \frac{1}{T} \int_0^T g(t,\epsilon) e^{i k \Omega t}
\label{eq:appfourier}
\end{align}

where $\Omega = 2\pi/T$ and $T$ is the Floquet period of the protocol. 

To compute the condutance, we need to evaluate the tunneling current. This can be evaluated at the edge of interest in the single lead experiment, or on either edge of the wire in the two lead setup. Without loss of generality, we evaluate the current at the left wire edge---

\begin{align}
I_L (t) &= -i e \left( \sum_q V_{L q} c^\dagger_{L q} (t) (t) c_{1} (t) - V^*_{L q} c^\dagger_{1} (t) c_{L q} (t) \right) \nonumber \\
&= -e \left(c^\dagger_1 (t) \xi_L (t) + \xi^\dagger_L (t) c_1 (t) \right) \nonumber \\
+& e\int_0^\infty d\tau \; \left[ \Gamma_L (\tau) c^\dagger_1 (t) c_1 (t - \tau) + \Gamma^*_L ( \tau) c^\dagger_1 (t- \tau) c_1 (t) \right]. 
\end{align}

Substituting the solution of the Heisenberg equation of motion, Eq.~(\ref{eq:app3}), and substituting thermal correlators for the lead operators, we find for the first and second terms

\begin{align}
I^{(1)}_L (t) &= - \frac{i e}{2 \pi} \int_{-\infty}^{\infty} d \epsilon \Gamma_L (\epsilon) f_L (\epsilon) \left[ G_{11} (t, \epsilon) - G^*_{11} (t, \epsilon) \right], \nonumber \\
I^{(2)}_L (t) &= \frac{e}{2 \pi} \int_0^\infty d \tau \int_{-\infty}^{\infty} d \epsilon \sum_l \bigg[ \Gamma_l (\tau) G^*_{1 l} (t, \epsilon) G_{1 l} (t - \tau, \epsilon)   \nonumber \\
&f_l (\epsilon) \Gamma_l (\epsilon) e^{i \epsilon \tau} + \Gamma_l (\tau) F_{1 l} (t, \epsilon) F^*_{1 l} (t-\tau, \epsilon) \left(1 - f_l (\epsilon) \right) \nonumber \\
&\Gamma_l (\epsilon e^{-i \epsilon \tau} \bigg] + \text{c.c.}. 
\label{eq:Iapp}
\end{align}

To massage these equations into final forms whose physical origin will be easy to surmise, it is useful to remove terms that contain `back-scattering' terms like $G_{11}$. For this, we use two equations whose derivation follows from similar discussion in Ref.~\cite{kohler2005driven}. We find 

\begin{align}
    G^\dagger (t, \epsilon) &- G(t, \epsilon) = i \frac{d}{dt} \;  G^\dagger (t, \epsilon) G(t, \epsilon) - G^\dagger (t, \epsilon) \Delta(t) F (t, \epsilon) \nonumber \\ 
    &+ i \int_0^\infty d \tau \; G^\dagger (t, \epsilon') \Gamma ( \tau) G (t - \tau, \epsilon) e^{i \epsilon t} + \text{h.c.}, \nonumber \\
    0 &= i \frac{d}{dt} \;  F^\dagger (t, \epsilon) F(t, \epsilon) + G^\dagger (t, \epsilon) \Delta(t) F (t, \epsilon) \nonumber \\
    &+i \int_0^\infty d \tau \; F^\dagger (t, \epsilon) \Gamma^*(\tau) F(t - \tau, \epsilon) e^{i \epsilon \tau} + \text{h.c.}. 
    \label{eq:app4}
\end{align}

Plugging in the sum of Eqs.~(\ref{eq:app4}) into Eqs.~(\ref{eq:Iapp}), using the Fourier decomposition of Eq.~(\ref{eq:appfourier}), and taking the time-average of the current over a Floquet period, we find our final result

\begin{align}
    \bar{I}_L &= \frac{e}{h} \sum_k \int_{-\infty}^{\infty} d \epsilon \;  f_R (\epsilon) \Gamma_R (\epsilon) \abs{G^{(k)}_{1N} (\epsilon)}^2 \Gamma_L ( \epsilon + k \Omega) \nonumber \\
    &+ \frac{e}{h} \sum_k \int_{-\infty}^{\infty} d \epsilon \; \left( 1- f_R (\epsilon) \right) \Gamma_R (\epsilon) \abs{F^{(k)}_{1N} (\epsilon)}^2 \Gamma_L (-\epsilon - k \Omega) \nonumber \\
    &- \frac{e}{h} \sum_k \int_{-\infty}^{\infty} d \epsilon \; f_L (\epsilon) \Gamma_L (\epsilon) \abs{G^{(k)}_{N1} (\epsilon)}^2 \Gamma_R ( \epsilon + k \Omega) \nonumber \\
    &- \frac{e}{h} \sum_k \int_{-\infty}^{\infty} d \epsilon \; f_L (\epsilon) \Gamma_L (\epsilon) \abs{F^{(k)}_{N1} (\epsilon)}^2 \Gamma_R ( -\epsilon - k \Omega) \nonumber \\
    &- \frac{e}{h} \sum_k \int_{-\infty}^{\infty} d \epsilon \; f_L (\epsilon) \Gamma_L (\epsilon) \abs{F^{(k)}_{11} (\epsilon)}^2 \Gamma_L (-\epsilon - k \Omega) \nonumber \\
    &+ \frac{e}{h} \sum_k \int_{-\infty}^{\infty} d \epsilon \; \left( 1 - f_L (\epsilon) \right) \Gamma_L (\epsilon) \abs{F^{(k)}_{11} (\epsilon)}^2 \Gamma_L (-\epsilon - k \Omega) \nonumber \\
    \label{eq:Ifinalapp}
\end{align}

The above terms can be easily interpreted as standard tunneling processes which give rise to a time-averaged current. In particular, the first linear can be understood as follows---it is a the rate at which an electron from the right lead can tunnel into the wire at energy $\epsilon$, $f_R (\epsilon) \Gamma_R (\epsilon)$, times the amplitude for traveling from site $N$ to site $1$ while absorbing $k$ photons of drive frequency $\Omega$, and then escaping to the left lead with rate $\Gamma_L (\epsilon + k \Omega)$ at the energy $\epsilon + k \Omega$. Similarly, the second term involves the tunneling of a hole from the right lead, and hence comes with the hole occupation factor,  $\left(1 - f_l (\epsilon) \right)$; the hole then subsqequently gets converted to an electron in the wire and emits $k$ photons, before eventually escaping to the left lead at the appropriate energy. 

Note importantly that lines 1-4 of Eq.~(\ref{eq:Ifinalapp}) vanish in the single lead problem we study and the only remaining terms describe Andreev reflections on the left edge of the wire. In what follows, we will describe the tunneling conductance assuming a connection only to the left lead. 

Finally, we assume a constant density of states for lead modes, thus $\Gamma_L (\epsilon)$ is independent of $\epsilon$. The particle-hole symmetry in the superconducting setting implies $\abs{F_{11} (\epsilon)}^2 = \abs{F_{11} (-\epsilon)}^2 $. At bias voltage $V$, the tunneling conductance reads 

\begin{align}
\sigma(V, T = 0) &= \frac{2e^2}{h} \sum_k \abs{F^{(k)}_{11} (\epsilon)}^2 \Gamma^2_L \nonumber \\
\sigma(V, T) &= \int d \epsilon \; \sigma( \epsilon, T = 0) \frac{\beta}{4 \text{cosh} \left(\frac{\beta (\epsilon + eV)}{2} \right)^2 } , 
\end{align}

The above expressions stand for the spinless case, but may be generalized immediately to the spinful case. First, note that although we've used indicies $n$ to quantify lattice sites, they may be used just as well to enumerate the electron spin. Next, we note that there are four matrix elements relevant for computing conductance in the spinful case---electrons tunneling in with spin up/down and exiting with up/down. We can simply add all these contributions to compute the net conductance.

\section{Conductance in the presence of MZMs}
\label{sec:condMZM}

We confirm here that the expression above gives the expected $2e^2/h$ conductance for MZMs in the static setting. A constant density of states implies $\Gamma (\tau) = \Gamma \delta (\tau)$. This is equivalent to a Markovian approximation for the memory function in the equation for the Green's function---

\begin{align}
    \d{g}{t} &= - i \delta (t - t') \mathbb{1} - i h g - \int_0^\infty d \tau \; d (\tau) g ( t - \tau) \nonumber \\
    &= - i \delta (t - t') \mathbb{1} - i h g - \frac{d}{2} g ( t - \tau), 
\end{align}

which yields $g (\epsilon) = \frac{1}{\epsilon - h + i \frac{d}{2}}$. 

Let's assume that the only modes that appear close to zero energy are the MZMs, at energies $\pm \epsilon_1$. In terms of these modes, represented by kets $\ket{\xi_1}, \ket{\bar{\xi}_1}$, the Hamiltonian and the dissipation matrix $d$ become

\begin{align}
    h &= \epsilon_1 \ket{\xi_1} \bra{\xi_1} - \epsilon_1 \ket{\bar{\xi}_1} \bra{\bar{\xi}_1}, \nonumber \\ 
    d &= \left( \ket{\xi_1} \bra{\xi_1} + \ket{\bar{\xi}_1} \bra{\bar{\xi}_1} \right) \Gamma \left(\abs{u_1}^2 + \abs{v_1}^2 \right) \nonumber \\
    &+ 2 \Gamma u_1 v_1 \ket{\bar{\xi}_1} \bra{\xi_1} + \text{h.c.}, 
\end{align}

where note that $u_1, v_1$ are the particle and hole amplitudes of the mode $\ket{\xi_1}$ on the first site on the wire (to which the lead couples to). Defining $B = u_1 v_1, A = \frac{1}{2} \left(\abs{u_1}^2 + \abs{v_1}^2 \right)$, we find

\be
F_{11} (V) = \frac{2 v^*_1 u_1 V}{V^2 - \epsilon^2_1 + 2 i \epsilon \Gamma A + \Gamma^2 \left( \abs{B}^2 - A^2 \right)}
\ee

At zero temperature, it is evident that the tunneling conductance (proportional to the above matrix element) at precisely zero bias is zero; of course, this is a purely zero temperature effect and the tunneling amplitude at zero bias is immediately filled in at finite temperature. Now, if one takes the limit $\epsilon_1 \rightarrow 0$ (perfect MZMs), this is accompanied by the symmetrization of the particle and hole components of the MZMs with $\abs{B} \rightarrow A$. Thus, for tunneling bias $V$, we find 

\begin{align}
\sigma (V) = \frac{2e^2}{h} \Gamma^2 \abs{F_{11} (V)}^2 = \frac{2e^2}{h} \frac{4 \Gamma^2 \abs{u_1}^2 \abs{v_1}^2 }{V^2 + \Gamma^2 (\abs{u_1}^2 + \abs{v_1}^2) } 
\end{align}

where in the limit of perfect MZMs, $\abs{u_1} = \abs{v_1} = 1/\sqrt{2}$, and we find $\sigma(0) = 2e^2/h$. 

\section{Numerical details of calculation of Green's functions for conductance}
\label{sec: numdetcond}

The Green's function equation, Eq.~(\ref{eq:gapp}) can be solved formally in terms of the evolution matrix $U(t,t') = T_t \left\{ e^{-i \int_{t'}^t d t \; h - i d/2} \right\} $, with $G(t,t') = -i \Theta ( t- t') U (t,t')$, and where $T_t \{ \cdot \}$ indicates time-ordering.

Bloch's theorem implies that one can write down the time-evolution matrix in terms of Floquet eigenstates $\ket{u_\alpha (t)}$ of the time-evolution matrix $U (T,0)$, with complex eigenenergy, $\epsilon_\alpha - i \gamma_\alpha$. We find these eigenstates by numerically evaluating and diagonalizing $U (T,0)$. In terms of these eigenstates, one may write 

\be
U_T (t,t') = \sum_\alpha e^{-i (\epsilon_\alpha - i \gamma_\alpha) (t - t') } \ket{u_\alpha (t)} \bra{u^+_\alpha (t')}
\ee

where $\bra{u^+_\alpha (t')}$ together with $\ket{u_\alpha (t)}$ form a bi-orthogonal basis satisfying $\braket{u^+_\alpha (t)}{u_\beta (t)} = \delta_{\alpha \beta}$. In practice, these orthogonal states are found simply by inverting the matrix of Floquet eigenvectors at all times. 

The Green's function can be evaluated finally as 

\begin{align}
G^{(k)} (\epsilon) &= -i \int_0^T \frac{dt}{T} e^{i k \Omega t}
\int_0^\infty d \tau \; e^{i \epsilon \tau} U (t, t-\tau) \nonumber \\
G^{(k)} (\epsilon) &= \sum_{\alpha, k'} \frac{\ket{u_{\alpha, k'+k}} \bra{u^+_{\alpha, k'}} }{\epsilon - \left( \epsilon_\alpha + k' \Omega - i \gamma_\alpha \right) }
\end{align}

\section{Details on tracking particle density on the QD}
\label{sec:detailsontracking}
Here we detail the expression for the number density on the QD that confirms the LZ transitions for the proposed protocols.  

The average number density on the QD for the realistic nano-wire setup at every time is defined as 
\begin{equation}
\begin{aligned}
\left\langle n_d(t)\right\rangle = &\left\langle n_{d\uparrow}(t)\right\rangle +\left\langle n_{d\downarrow}(t)\right\rangle\\
=&\bra{\psi_w}\otimes\bra{1_{d\uparrow}}c^\dagger_{d\uparrow}(t)c_{d\uparrow}(t)\ket{1_{d\uparrow}}\otimes\ket{\psi_w}\\
+&\bra{\psi_w}\otimes\bra{1_{d\uparrow}}c^\dagger_{d\downarrow}(t)c_{d\downarrow}(t)\ket{1_{d\uparrow}}\otimes\ket{\psi_w}
\end{aligned}
\end{equation}
where $\ket{1_{d\uparrow}}$ is a spin-up occupied QD state which represents the initial state of the QD and $\ket{\psi_w}$ represents an arbitrary wire state which we average over later. We can expand the time dependent fermion creation/annihilation operators for both spins in the basis of lattice fermion operators as

\begin{align}
\label{eq:d2}
c_{d\uparrow}(t)&=\sum^L_{n=d}\left[u^*_{n\uparrow}(t)c_{n\uparrow}+v^*_{n\uparrow}(t)c^\dagger_{n\uparrow}+u^*_{n\downarrow}(t)c_{n\downarrow}+v^*_{n\downarrow}(t)c^\dagger_{n\downarrow}\right] \nonumber \\
c_{d\downarrow}(t)&=\sum^L_{n=d}\left[w^*_{n\uparrow}(t)c_{n\uparrow}+z^*_{n\uparrow}(t)c^\dagger_{n\uparrow}+w^*_{n\downarrow}(t)c_{n\downarrow}+z^*_{n\downarrow}(t)c^\dagger_{n\downarrow}\right]
\end{align}

Upon averaging over arbitrary wire states, setting $\bra{\psi_w}c^\dagger_{n\uparrow}c_{n\uparrow}\ket{\psi_w} =  \bra{\psi_w}c^\dagger_{n\downarrow}c_{n\downarrow}\ket{\psi_w} = 1/2$, we find  
\begin{align}
\label{eq:updensity}
\left\langle n_{d\uparrow}(t)\right\rangle &=
    |u_{d\uparrow}(t)|^2+|v_{d\downarrow}(t)|^2 \nonumber \\
    &+\frac{1}{2}\sum^L_{n=1}\left(|u_{n\uparrow}(t)|^2+|v_{n\uparrow}(t)|^2+|u_{n\downarrow}(t)|^2+|v_{n\downarrow}(t)|^2\right). \nonumber \\
\left\langle n_{d\downarrow}(t)\right\rangle &=
    |w_{d\uparrow}(t)|^2+|z_{d\downarrow}(t)|^2 \nonumber \\
    &+\frac{1}{2}\sum^L_{n=1}\left(|w_{n\uparrow}(t)|^2+|z_{n\uparrow}(t)|^2+|w_{n\downarrow}(t)|^2+|z_{n\downarrow}(t)|^2\right).
\end{align}

We can further simplify the above expressions by noting that $\{c_{d\sigma}(t),c^\dagger_{d\sigma}(t)\}=1$, for all time, which results in a normalization condition on the coefficients in Eq.~\ref{eq:d2}. This leads to the result

\begin{align}
\label{eq:ndspinful}
    \left\langle n_{d\uparrow}(t)\right\rangle &=\frac{1}{2}+\frac{1}{2}\left(|u_{d\uparrow}(t)|^2+|v_{d\downarrow}(t)|^2-|u_{d\downarrow}(t)|^2-|v_{d\uparrow}(t)|^2\right)  \nonumber \\     \left\langle n_{d\downarrow}(t)\right\rangle &=\frac{1}{2}+\frac{1}{2}\left(|w_{d\uparrow}(t)|^2+|z_{d\downarrow}(t)|^2-|w_{d\downarrow}(t)|^2-|z_{d\uparrow}(t)|^2\right)
\end{align}
For the Kitaev chain, we can infer the expression for $\left\langle n_d(t)  \right\rangle$ by getting rid of the spin degrees of freedom from the realistic nano-wire case and conclude
\begin{equation}
\label{eq:nd}
\left\langle n_d(t)  \right\rangle = \frac{1}{2}+\frac{1}{2}\left(|u_d(t)|^2-|v_d(t)|^2 \right).
\end{equation}
We employ \eqref{eq:ndspinful} and \eqref{eq:nd} to numerically track the particle density on the QD site at every time step to confirm ideal LZ transitions.  

\section{Deviation from standard LZ process}
\label{sec:deviationLZ}

Here we note that for certain large values of $t_0$, the tunneling amplitude between the quantum dot and the wire, the usual avoided level crossing picture of the Landau Zener transitions gets distorted and results in a consecutive Landau Zener transitions. Although this does not change our conclusions drastically, it can lead to non-perturbative effects, including resonance effects that are seen in Landau-Zener-Stuckelberg based interferometry. This may be one potential cause of non-monotonic changes to the Landau-Zener tunneling probability seen in our numerical data. 

\begin{figure}[htp]
\includegraphics[width=3.2in]{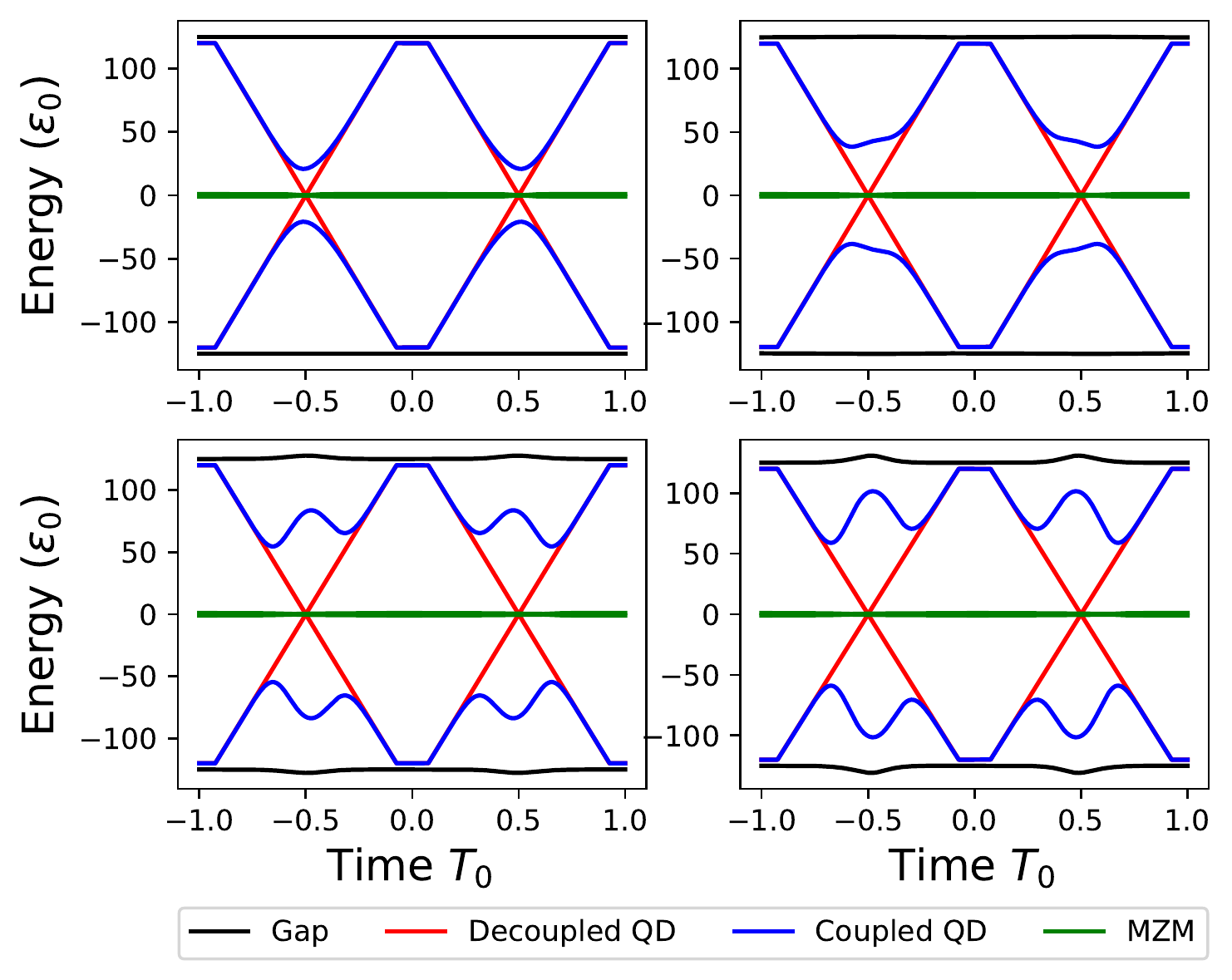}
\caption{Low energy instantaneous eigenenergies for the Parity Flip protocol in the QD+Kitaev chain at $\mu=1.8$ at $t_0=0.2,0.4,0.8$, and $1.0$ in units of $2\epsilon_\text{max}$. For small coupling $(t_0=0.2)$, we see the standard LZ picture where a simple gap with single bump emerge. We quickly deviate from this simple picture when the coupling is increased and the standard probability that describes non-adiabatic transition does not directly apply.}
\label{fig:instantneouseig}
\end{figure}

\bibliographystyle{apsrev4-1}
\bibliography{BraidingNew2}

\end{document}